\documentclass[preprint2]{aastex} 
\usepackage{graphicx} 
\usepackage{subfigure}
\usepackage{upgreek}
\usepackage{amssymb,amsmath}

\def\widefield{wide-field}
\def\Widefield{Wide-field}

\def\fig{Figure}

\def\Fig{Figure}

\def\sect{Section}

\def\Sect{Section}

\def\tab{Table}

\def\Tab{Table}

\def\I{~\textsc{i}}
\def\II{~\textsc{ii}}
\def\III{~\textsc{iii}}

\def\blobcat{\texttt{BLOBCAT}}
\def\emerlin{\textit{e}-MERLIN}

\def \rah   {\hbox{$^{\rm h}$}}
\def \ram   {\hbox{$^{\rm m}$}}
\def \ras   {\hbox{$^{\rm s}$}}

\begin{document}
\shorttitle{\Widefield\ VLBI observations of M31}
\shortauthors{Morgan et al.}
\title{\Widefield\ VLBI observations of M31: A Unique Probe of the Ionized ISM of a nearby galaxy}
\author{John S. Morgan\altaffilmark{1}, Megan K. Argo\altaffilmark{1, 2}, Cathryn M. Trott\altaffilmark{1, 3}, Jean-Pierre Macquart\altaffilmark{1}, Adam Deller\altaffilmark{2}, Enno Middelberg\altaffilmark{4}, James Miller-Jones\altaffilmark{1}, Steven J. Tingay\altaffilmark{1}} 
\email{john.morgan@icrar.org}
\altaffiltext{1}{International Center for Radio Astronomy Research, Curtin University, GPO Box U1987, Perth, WA 6845, Australia.}
\altaffiltext{2}{ASTRON, Oude Hoogeveensedijk 4, 7991 PD Dwingeloo, The Netherlands.}
\altaffiltext{3}{ARC Centre of Excellence for All-Sky Astrophysics.}
\altaffiltext{4}{Astronomisches Institut der Ruhr-Universit\"at Bochum, Universit\"atsstra\ss e 150, 44801 Bochum, Germany.}

\begin{abstract}
The VLBA was used at 1.6\,GHz to observe a target field 50\arcmin\ in diameter including the core of M31.
Novel VLBI correlation techniques were used to observe 200 sources simultaneously, of which 
16 were detected.
We classify all 16 as background AGN based on their X-ray properties and arcsecond- and mas-scale morphology.
The detected sources were then analyzed for evidence of scatter-broadening due to the ionized ISM of M31.
The detection of a compact background source only 0.25\,kpc projected distance from M31 places a constraint on the extent of any extreme scattering region associated with center of M31. 
However, the two sources closest to the core show evidence of scatter broadening consistent with that which would be observed if a compact source were observed through the inner disk of our Galaxy at the inclination of M31.
We interpret this as a detection of the ionized ISM of M31 along two lines of sight.
With the increases in bandwidth and sensitivity envisaged for future long-baseline interferometers, this should prove to be a remarkably powerful technique for understanding the ionized ISM in external galaxies.
\end{abstract}

\section{Introduction}
\Widefield\ Very Long Baseline Interferometry (VLBI) is a technique which allows simultaneous observations of multiple sources in a single pointing.
One of the main applications of \widefield\ VLBI since its earliest development \citep{Pedlar:1999} has been the study of nearby galaxies, particularly starburst galaxies such as M82 \citep{Fenech:2010}, NGC253 \citep{Lenc:2006}, and NGC4945 \citep{Lenc:2009}.
Such studies have provided insights into a wealth of astrophysical processes, providing estimates of the supernova rate in these galaxies, and the electron density of the ionized medium that is giving rise to the starbust.

M31 has not, until now, been the subject of such a study; we ascribe two reasons for this.
Firstly, the large angular extent of the galaxy means that it is only with recent developments in VLBI correlation and data reduction techniques \citep{Deller:2011,Morgan:2011} that it has been possible to survey any more than the tiniest fraction of M31 using VLBI.
Secondly it is not a starburst galaxy and so is unlikely to have the large population of young Supernova Remnants (SNRs) and exotic objects present in these more extreme environments.

However, with the increasing sensitivity of VLBI due to bandwidth upgrades such as that currently underway at the VLBA\footnote{Very Long Baseline Array}, compact sources intrinsic to M31 may be observable.
For example at the distance of M31, the Pulsar Wind Nebula \citep[PWN,][]{Gaensler:2006} associated with the Crab pulsar would have a total flux of 6\,mJy and an extent of approximately 1\arcsec\ with filamentary structure on a scale of 15\,mas \citep{Bietenholz:1997}, well matched to the resolution and sensitivity of 1--2\,GHz VLBA observations.
There are also objects similar to PWNe but which are powered by microquasars, for example the W50 nebula associated with SS443 \citep{Dubner:1998}. An object such as this would also be detectable in our observations.
In addition, an ultra-luminous stellar-mass microquasar within our target field was reported by \citet{Middleton:2013}.
\Widefield\ VLBI studies of M31 therefore have potential to detect analogs of rare Galactic objects in other galaxies.

VLBI measurements of the proper motion of M31 are currently lacking, in contrast to many other nearby galaxies such as M33 \citep{Brunthaler:2005} and IC10 \citep{Brunthaler:2007}.
Deriving such proper motions via VLBI techniques requires both compact sources intrinsic to the galaxy (usually, though not necessarily, masers), and background compact sources which can be used as calibrators.
Discovery of the latter is yet to be reported.
Such proper motion measurements would be a useful complement to the recent proper motion for several fields observed at multiple epochs with the HST as reported by \citet{Sohn:2012}.

\subsection{M31}
\begin{figure}[t]
	\centering
	\includegraphics[clip, width=\columnwidth, trim=0 49 0 49 ]{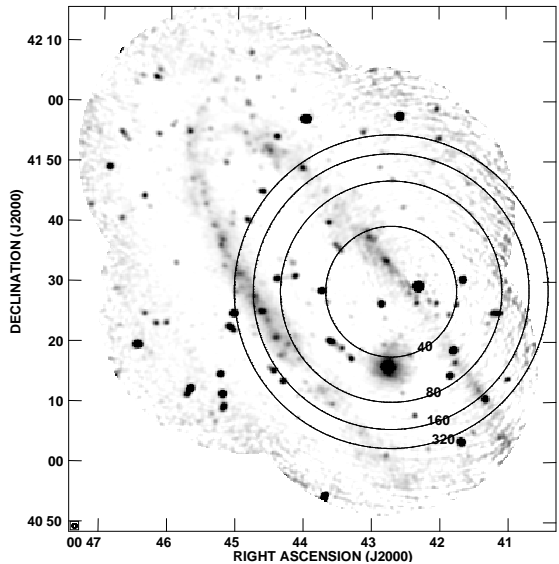}
	\figcaption{\label{fig:ba097_contours} Grayscale: A subset of data presented in \citet{Beck:1998}, extracted from the VLA archive. The core and 10\,kpc ring are clearly visible. Contours: predicted RMS sensitivity of our observation ($\upmu$Jy). The peak sensitivity is 30$\upmu$Jy RMS}
\end{figure}

M31 is the nearest non-dwarf neighbor to the Milky Way, at a distance of 785\,kpc \citep{McConnachie:2005}.
Its D25 isophote is an ellipse with major and minor axes of 155\arcmin\ and 50\arcmin\ respectively, and the inclination of the disk to our line of sight is approximately 77\degr\ \citep{Tully:1988}.
In all continuum surveys and in far-infrared images, diffuse emission from a central region some 4\arcmin\ (approximately 1\,kpc) in extent is seen.
The other notable feature both in low-resolution radio maps and in the far infrared is a ring some 10\,kpc distant from the center (see \fig~\ref{fig:ba097_contours}).
A rather less obvious but nonetheless important feature of M31 was revealed in H$\upalpha$, N\II\ and O\III\ imaging of the core carried out by \citet{Jacoby:1985}.
They revealed a striking spiral structure (with an entirely different orientation to the main galactic plane) in the ionized gas occupying this central region.

M31* is the M31 analog of Sgr~A*: an X-ray and radio source corresponding to a supermassive black hole of mass $1.4^{+0.7}_{-0.3}\times10^8\mathrm{M}_{\odot}$ \citep{Bender:2005} (c.f. Sgr~A* $\sim4.0\times10^6\mathrm{M}_{\odot}$).
It is three times less luminous than Sgr~A* in the radio and three times more luminous in X-rays \citep{Garcia:2010}, significantly below the fundamental plane.

The estimated star formation rate in M31 is significantly less than in nearby starburst galaxies.
\citet{Williams:2003} found the optical mean star formation rate for the disk of M31 to be $\sim$1\,M$_{\odot}$\,yr$^{-1}$, approximately half that found in M82 from radio observations; while \citet{Kang:2009} found an average of 0.6 to 0.7\,M$_{\odot}$\,yr$^{-1}$ over the last 400\,Myr from UV observations.
One supernova has been observed in M31, very close to the center \citep{Hartwig:1885}; and a number of SNRs in the core have been identified via their X-ray properties \citep{Kong:2003}, and throughout the North-Eastern half of the galaxy via their forbidden line characteristics \citep{Braun:1993}.

Unsurprisingly, there exists a great deal of archival data on M31.
Radio continuum surveys are summarized in \tab~\ref{table:surveys}. The highest resolution survey to date is that of \citet[][hereafter B90]{Braun:1990} which imaged just over one half of M31 using 10 pointings of the VLA\footnote{Very Large Array (now upgraded and renamed the Karl G. Jansky Very Large Array)} at 1.4\,GHz with a resolution of 5\arcsec\ using the B-, C-, and D-configurations.
This survey detailed 534 individual continuum sources, a large fraction of which were unresolved and of which 103 have possible associations with H$\upalpha$ nebulae and seven with previously identified SNRs.

The resulting map also contains 121 objects associated with previously known radio sources detected in WSRT\footnote{The Westerbork Synthesis Radio Telescope} surveys of the entire galaxy at 610\,MHz \citep{Bystedt:1984} and at 1412\,MHz \citep{Walterbos:1985}, the 36W and 37W catalogs respectively.
In addition, \citet{Beck:1998} surveyed the entire galaxy at 20\,cm using both the Effelsberg dish and the VLA in the compact D configuration, concentrating on polarization properties of the M31 field.
Subsequently the entire M31 field was surveyed by \citet{Gelfand:2004} (hereafter GLG) with a single pointing of the VLA in A-configuration at 325\,MHz, uncovering a total of 405 sources across 7.6\,deg$^2$.

In the X-ray, \citet{Stiele:2011} used archival data, supplemented with new observations to produce a deep XMM-Newton Survey of the entire disk, producing a catalog of 1987 sources.

In H\I\ emission studies, the state-of-the-art is the extensive study of \citet{Braun:2009}.
Surveys of H\II\ regions have been undertaken both by \citet{Walterbos:1992}  and \citet{Azimlu:2011}.

\begin{deluxetable}{l l l r r l l}
	\tablecolumns{7}
	\tablewidth{0pc}
	\tableheadfrac{0.05}
	\tabletypesize{\scriptsize}
	\tablecaption{Radio Continuum Surveys of M31.\label{table:surveys}}
	\tablehead{
	\colhead{Survey} 	& \colhead{{Telescope}} & \colhead{Frequency}	&\colhead{Resolution} 	& \multicolumn{2}{c}{RMS Sensitivity}	& \colhead{Reference} \\
\colhead{}	 	& \colhead{}		& \colhead{(GHz)}	&\colhead{(\arcsec)} 	& \colhead{(mJy)} 		& \colhead{} \\ }
\startdata
B90 		&  VLA-B,C,D	&  1.5		&  5		&  0.03 & uniform	& \citet{Braun:1990} \\
37W		&  WSRT 	&  1.4		&  23$\times$35	&  0.2 	& peak		& \citet{Walterbos:1985} \\ 
36W		&  WSRT 	&  0.61 	&  54$\times$82	&  0.8	& peak		& \citet{Bystedt:1984} \\ 
GLG		&  VLA-A	&  0.325	&  6		&  0.6 	& peak		& \citet{Gelfand:2004} \\
NVSS		&  VLA-D	&  1.4		&  40		&  0.45 & uniform	& \citet{nvss:1998} \\
WENSS		&  WSRT		&  0.325	&  54$\times$82	&  3.6	& uniform	& \citet{Rengelink:1997} \\
VLSS		&  VLA-A	&  0.070	&  80    	&  100	& uniform	& \citet{Cohen:2007} \\
\enddata
	\tablecomments{All but B90 cover the entire disk.}
\end{deluxetable}

We have undertaken a pilot study to explore a field of M31 with \widefield\ VLBI, in detail, for the first time.
In this paper we present these observations, supplemented by some data from the VLBA archive and VLA observations.
In \sect~\ref{sec:observations} we describe the observations and our source selection and data reduction procedure; 
in \sect~\ref{sec:results} we describe the results of our observations and cross-match our sources with previous surveys and data from the VLBA archive;
in \sect~\ref{sec:discussion} we discuss our results.

\section{Observations}
\label{sec:observations}
An 8-hour observation was carried out with all 10 antennas of the VLBA on 2010-July-04.
The target field was centered on 00\rah42\ram43.3\ras, 41\degr28\arcmin25\arcsec\ (J2000.0).
It was chosen using the following criteria: 1) overlap with the B90 survey; 2) overlap with both the core and a segment of the starforming ring; 3) inclusion of bright, unresolved sources which provide a high probability of easy detection (\fig~\ref{fig:ba097_contours}).
The source \object{ICRF J003824.8+413706} was used as a phase reference calibrator, with a duty cycle of 5 minutes on target, 1 minute on calibrator.
The source \object{3C84} was used as a fringe finder and bandpass calibrator and was  observed for 5 minute scans 3 times during the observation.

64\,MHz of contiguous bandwidth between 1608.49 and 1672.49\,MHz was recorded in 4$\times$16 MHz sub-bands.
Both polarizations were recorded with 2-bit precision for a total data rate of 512\,Mbps per antenna.

\subsection{Target Selection and Correlation}
\label{sec:correlation}
Correlation was carried out using the DiFX software correlator \citep{Deller:2007}.
The multi-phase center technique \citep{Deller:2011} was vital for correlation of the target field as this allowed multiple phase centers to be correlated simultaneously in a single pass.

The remarkable efficiency of the DiFX correlator in multi-phase center mode means that there is little processing time to be saved in being economical with the number of phase centers.
Consequently, we did not consider whether a source showed a compact morphology, or if it was bright enough to be detected (while the sensitivity of our observations equal that of the B90 survey at our pointing center, primary beam effects mean that this is certainly not the case over the entire field).
This was justified on the basis that any source might be variable, and any resolved source could have a compact component and/or be masking a separate compact source.

All B90 sources closer than 20\arcmin\ to the field center were targeted.
Use was also made of the 37W catalog, which covers a wider area and also gives separate coordinates for each source component: all 37W source components within 30\arcmin\ were targeted.
In addition, sensitive VLA images of the core region of M31 were extracted from the NVAS archive \citep{Crossley:2007}, however no firm detections could be discerned other than those already included in the B90 catalog.
Any source which was within 10\arcsec\ of another was removed from the list of correlation targets.
The output of the correlator consisted of 195 separate visibility datasets, each phase-centered on one of our target sources.

\subsection{VLBI Data Calibration}
For the most part, calibration of a \widefield\ VLBI dataset proceeds in an identical fashion to a standard VLBI observation.
First, extensive flagging was carried out via visual inspections of the visibilities.
Next, corrections to the amplitude were made using the AIPS task `ACCOR'.
Our own ParselTongue \citep{Kettenis:2006} scripts were used to apply phase corrections based on pulse cal tone information extracted by DiFX.
DiFX extracts all pulse cal tones from across the band, rather than just the edge tones, increasing the distance between 2$\pi$ ambiguities of phase jumps to 1$\upmu$s.
Visual inspection of the visibilities confirmed that this ensured that all sub-bands were coherent with each other throughout the observation (in contrast to the standard VLBA procedures, which were unable to correctly adjust for phase jumps of $>0.07\upmu$s).

Next, system temperature measurements and gain curves were used to provide amplitude calibration using the AIPS tasks `APCAL'.
Online system temperature measurements for the Hancock antenna were found to be invalid and so a nominal value of 40\,K was used throughout.

A combined fringe fit solution for all sub-bands was then derived using the AIPS task `FRING', assuming a point source calibrator and a fringe rate of zero.
The resulting calibration solution was then applied for a first imaging attempt.

\subsection{\Widefield\ Data Calibration and Imaging}
\label{sec:wfdatareduction}
Before applying any calibration, phase centers for the `duplicate' sources removed from the correlation target list (see \sect~\ref{sec:correlation}) were regenerated using the AIPS task `UVFIX' to shift the phase center of the nearest correlated phase center to this point.
This resulted in 226 visibility datasets consisting of 147 B90 sources and 86 components of 79 sources from the 37W catalog.

The calibration solution as derived in the previous section was then applied to all these sources.
In addition primary beam corrections as described by \citet{Middelberg:2013} were applied.
In this scheme, the primary beam is assumed not to vary across the area which is to be imaged, allowing the primary beam corrections to be applied directly to the visibilities, taking into account the slight (but significant) dependence of the correction on antenna, polarization, and hour angle.
The frequency scaling of the primary beam was calculated for each sub-band.
A primary beam model as described by \citet{Uson:2008} was used with parameters measured for each individual VLBA antenna (R.C. Walker Priv. Comm.; see also \citeauthor{Middelberg:2013} \tab~3).
Such corrections were verified by \citeauthor{Middelberg:2013} to be accurate at a level of 4\%.

Preliminary images of all target sources were then produced.
The RMS noise level in the image was determined using the AIPS task `IMEAN'.
Comparison of this value to the image peak value was sufficient to easily identify the brighter of the detected target sources.

\subsection{Refinement of the Calibration}
For both datasets the coherence time was far greater than the phase reference duty cycle for the vast majority of the observation, given the close proximity of the calibrator to the target fields and the favorable time of the observation in the solar cycle.

Further careful flagging was carried out, including flagging all data whose $u$ coordinate was $<50\,{\rm k}\lambda$.
Such visibilities are susceptible to RFI due to the low fringe rate and this flagging was successful in removing horizontal stripes that had been seen in some images.

The detection of bright sources within the target field allows the possibility of using these sources as in-beam calibrators.
With sufficient signal to noise (S/N), these sources can be expected to give improved calibration solutions due both to their proximity to the other targets and the fact that they are observed simultaneously rather than in interleaved pointings.
One option was to use the multi-source phase referencing technique, as described in \citet{Middelberg:2013}, however the S/N of all our sources combined in quadrature would only be approximately 10\% greater than for the brightest source alone, which was deemed unlikely to lead to a significant improvement.

The use of 37W144 as an in-beam calibrator (phase only, with one combined solution for all sub-bands and polarizations) resulted in an average increase of S/N of most of the target sources of 10\%.

This refined calibration was then applied to all target visibility datasets and the data were then averaged in frequency to 8 spectral channels per sub-band.
These averaged data were used for all subsequent analysis.
Source detection images were then produced for all fields consisting of `dirty', normally-weighted maps.

\subsection{Source Detection}
Since the difference in synthesized beam size between the B90 observations and our VLBA observations is approximately 1000, a minimum of 1\,million resolution units must be searched for each source.
Imaging requirements meant that in reality 8192$\times$8192 pixel images with each pixel measuring 1\,mas on a side were used to image the 5\arcsec\ resolution unit of the B90 catalog.
Assuming a Gaussian distribution of pixel brightness, we would expect one 5.6\,$\sigma$ noise peak in every VLBI image.
We are therefore motivated to determine the probability that any apparent source is real, rather than just a noise spike.

We calculate a likelihood ratio (LR) for each pixel in each of our images formed from the following hypotheses:
1) the observed flux can be associated with a cataloged source ($H_1$);
2) the observed flux is a noise spike ($H_0$).
This calculation is made under the following assumptions:
1) the noise in each image is Gaussian-distributed;
2) the RMS of the noise does not vary across any individual VLBI image.

$H_0$ is simply the Cumulative Distribution Function of the positive half of the Gaussian distribution:
\begin{eqnarray}
p(m, \sigma_{\rm rms}; H_0) &=& {\rm CDF}_\mathcal{N}(m,\sigma_{\rm rms})\\
&=& \frac{1}{2}\left[ 1 - {\rm erf}{\left( \frac{m}{\sqrt{2\sigma_{\rm rms}^2}} \right)} \right],
\end{eqnarray}
where $m$ is the brightness of the pixel in the VLBI image and $\sigma_{\rm rms}$ is the RMS noise in the VLBI image.

$H_1$ is based solely on the position of the VLBI pixel and the cataloged source position/morphology.
\begin{eqnarray}
p({\bf r}, {\boldsymbol \theta}; H_1) &=& l({\bf r})\\
&=&  \varphi({\bf r}, \sigma_a, \sigma_b, \phi), \label{eqn:elliptical_gauss}
\end{eqnarray}
where ${\bf r}$ is the relative vector offset between cataloged position and VLBI pixel position.
The cataloged source structure ${\boldsymbol \theta}$ is described by $\sigma_a$, $\sigma_b$ and $\phi$ where these are the cataloged semi-major axis, semi-minor axis and position angle of the source respectively.
$l(r)$ is then calculated using a normalized elliptical Gaussian, $\varphi$, parameterized by these variables.
The instrumental resolution is also accounted for here as detailed below.

The Likelihood ratio (LR) is then simply
\begin{equation}
	{\rm LR} = \frac{p(H_1)}{p(H_0)}.
\end{equation}

A further refinement we considered was taking into account how well-matched the cataloged flux and the VLBI flux were.
For example, the catalog/VLBI flux scaling could have been modeled as a top-hat function convolved with a softening that incorporated image noise and variability softening.
This was not done for several reasons.
Firstly, any source could be variable---particularly on the 20-year timescale that separates our observations from the catalog observations.
Secondly, we expect that a significant fraction of the flux could be resolved out by the VLBA.
This leaves only cases where the VLBI detection is significantly brighter than its cataloged counterpart.
Since the object of this exercise is to discriminate between noise spikes and weaker sources, such a refinement would serve no purpose.

\subsubsection{Generating LR Maps}
While the quoted resolution for the B90 observations was 5\arcsec\, the astrometric precision could be expected to be greater for brighter sources.
For all B90 sources, $\sigma_a$ and $\sigma_b$ were set either to their corresponding values, or to the quoted error on these values, whichever was larger.

Likewise, \citet{Walterbos:1985} gave a major axis, minor axis, and a positional angle $\phi$ for each resolved source and an error in RA and declination for all sources.
The maximum values were used for $\sigma_a$ and $\sigma_b$.

FITS images corresponding to the VLBI images containing the LR values for each pixel were then generated.
Before any use was made of the VLBI maps, the outer 512 pixels from the edge were excluded from the analysis to mitigate the effect of aliasing effects.

\subsubsection{Determining the Detection Threshold}
\Tab~\ref{table:lr} gives the VLBI map statistics along with maximum value from the LR maps for all detected sources and the field containing the next highest LR pixel.

We note that the simple statistic of VLBI S/N is as good a predictor of whether a source is real as the likelihood ratio.
This reflects the extremely poor constraint on position provided by the catalogs, compared with the VLBI position.

These data show also that our assumption of pure Gaussian noise is a good one.
Our brightest noise spike in $\sim170\times7168\times7168$ pixels is 6.15\,$\sigma$.
A pixel of this value or brighter would be expected to occur in Gaussian noise once in every 50 maps.
Compare this with our weakest detection which would only occur in Gaussian noise once in every 64\,000 maps.
It should be stressed that such pure Gaussian noise in the non-detection maps was only achievable with very careful flagging of the visibility data and with masking of the VLBI images to mitigate aliasing effects at the image edge.

This very clear cutoff between our weakest detection and our brightest noise peak was confirmed upon visual inspection of the VLBI images and LR maps.
We note, however, that real sources, weaker than 6.15\,$\sigma$ almost certainly exist in our data, particularly in those fields near the edge of the primary beam, however we find no statistical evidence for them in this analysis. 

For the upper limit on the brightness of any non-detection, we adopt 6.6\,$\sigma$.
This corresponds approximately to P$<0.05$ of a source being a noise spike over $\sim200$ images.

\begin{deluxetable}{r r r r r r r r r}
	\tablecolumns{8}
	\tablewidth{0pc}
	\tableheadfrac{0.05}
	\tabletypesize{\scriptsize}
	\tablecaption{Detection Statistics\label{table:lr}}
	\tablehead{
	\multicolumn{5}{c}{All baselines} && \multicolumn{3}{c}{10\,M$\lambda$ cutoff} \\
	\cline{1-5}\cline{7-9}\\

		\colhead{$m_{\rm peak}$} &
		\colhead{$\sigma_{\rm rms}$} &
		\colhead{S/N } &
		\colhead{-$\log_{10}\left( p(H_0) \right)$} &
		\colhead{$\log_{10}\left(\rm LR\right)$} &&
		\colhead{S/N} &
		\colhead{-$\log_{10}\left( p(H_0) \right)$} &
		\colhead{$\log_{10}\left(\rm LR\right)$} \\

		\colhead{(mJy)} &
		\colhead{($\mu$Jy)} &
		\colhead{($m_{\rm peak}/\sigma_{\rm rms}$)} & & && & & \\
		}
	\startdata

11.95		&65			&180					&$>$308					&$>$308 			&&240				&$>$308					&$>$308 			\\  
4.64		&59			&79					&$>$308  				&$>$308 			&&64				&$>$308  				&$>$308 			\\  
4.70		&79			&60					&$>$308 				&$>$308 			&&77				&$>$308 				&$>$308 			\\  
3.14		&64			&49					&$>$308 				&$>$308 			&&41				&$>$308 				&$>$308 			\\  
0.73		&34			&21					&102     				&102    			&&22				&111     				&110    			\\  
1.64		&80			&20					&92.6   				&92.0   			&&15				&50.0   				&49.3   			\\  
2.35		&119			&20					&85.7   				&85.2   			&&13				&39.1   				&38.5   			\\  
1.96		&102			&19					&81.6   				&80.7   			&&14				&45.1   				&44.1   			\\  
17.46		&1074			&16					&58.9 					&56.1   			&&10				&26.0 					&23.2   			\\  
0.75		&58			&13					&37.6    				&37.0   			&&8.5				&16.9    				&16.4   			\\  
0.68		&53			&13					&37.3    				&36.8   			&&14 				&46.4    				&45.9   			\\  
2.68		&272			&9.8					&22.4   				&22.0   			&&8.0				&15.3   				&14.9   			\\  
0.50		&53			&9.5 					&20.8    				&20.3   			&&6.5				&10.3    				&9.9    			\\  
0.38		&42			&9.0					&19.1    				&18.7   			&&6.9				&11.5    				&11.1	  			\\  
0.53		&62			&8.5 					&17.2   				&16.7   			&&9.6 				&21.6   				&21.1   			\\  
0.40		&55			&7.2					&12.5    				&11.4   			&&8.8				&18.2    				&17.7				\\  
\hline                                                                                                                                                  
0.53		&87			&6.2 					&9.41   				&8.94   			&&6.3 				&9.81   				&9.37   			\\  
\enddata
	\tablecomments{The 16 firm detections are shown, along with the non-detection with the highest LR (below the horizontal line). These are different fields for the two different resolutions.  $m_{\rm peak}$, $p(H_0)$, and LR are shown for the brightest pixel in the image.}
\end{deluxetable}

\subsection{\Widefield\ Mapping}
In addition, each targeted source was imaged with visibilities on baselines longer than 10\,M$\lambda$ discarded (approximately corresponding to the inner 8/45 baselines).
This allowed the pixel size to be increased to 5\,mas with a commensurate increase in field size to a 40\arcsec\ diameter.
The purpose of this exercise was twofold: first, to detect any emission from extended parts of the sources; second, to search for less compact sources which would not be detected on the longer baselines.

The same likelihood analysis was then applied to these images.
No new sources were detected.
Though the S/N of some of the more compact sources dropped, the sources with the highest LR values were the 16 sources already detected (see \tab~\ref{table:lr}).

\subsection{Imaging and UV Modeling of Detected Sources}
\label{sec:modeling}
\begin{figure*}[p]
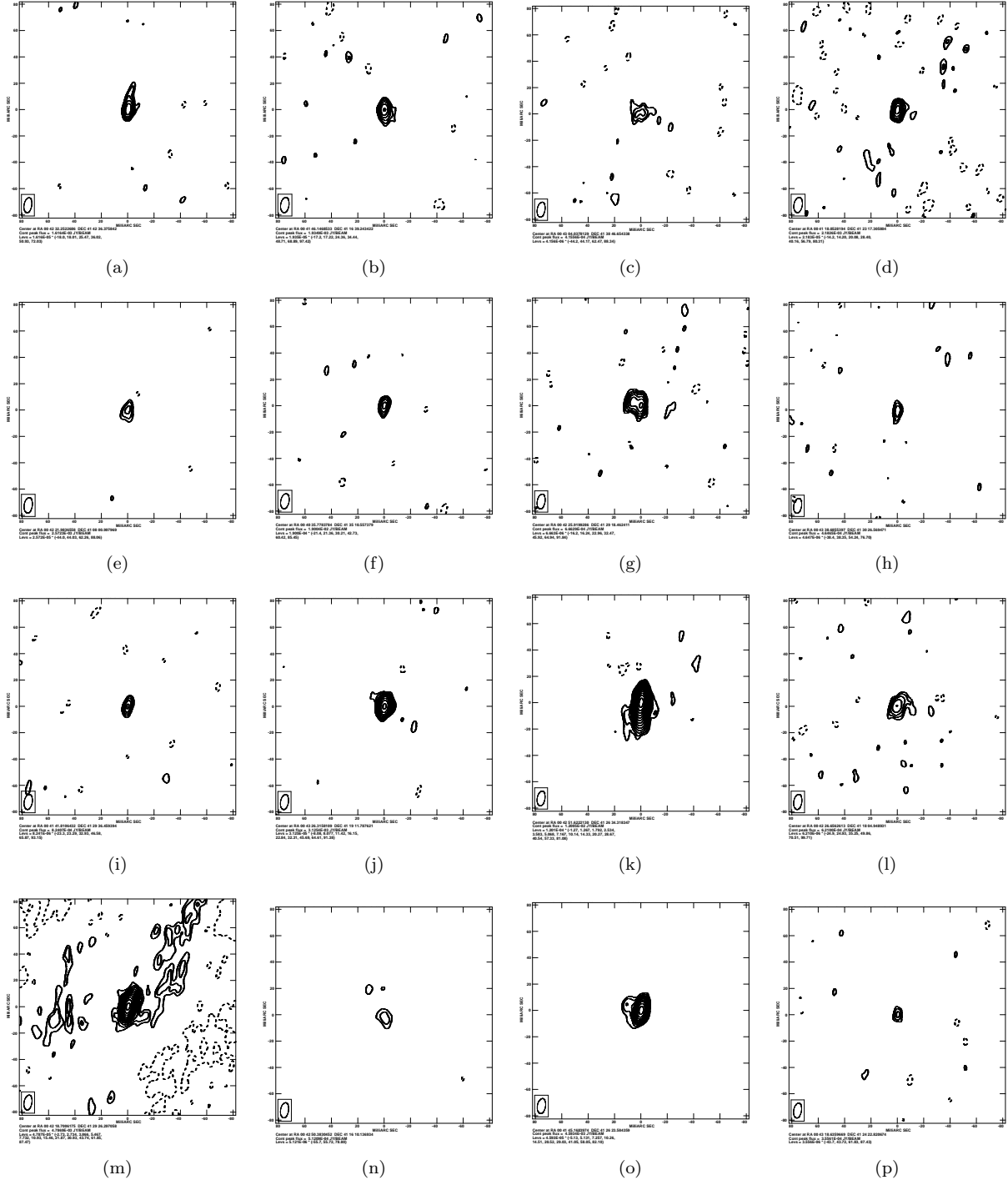

	\centering
	\subfigure[]{\label{fig:0099_im}\includegraphics[width=0.23\textwidth, clip]{0099_im.eps}} \, 
	\subfigure[]{\label{fig:0111_im}\includegraphics[width=0.23\textwidth, clip]{0111_im.eps}} \, 
	\subfigure[]{\label{fig:0058_im}\includegraphics[width=0.23\textwidth, clip]{0058_im.eps}} \, 
	\subfigure[]{\label{fig:0117_im}\includegraphics[width=0.23\textwidth, clip]{0117_im.eps}} \\ 
	\subfigure[]{\label{fig:0154_im}\includegraphics[width=0.23\textwidth, clip]{0154_im.eps}} \, 
	\subfigure[]{\label{fig:0167_im}\includegraphics[width=0.23\textwidth, clip]{0167_im.eps}} \, 
	\subfigure[]{\label{fig:0007_im}\includegraphics[width=0.23\textwidth, clip]{0007_im.eps}} \, 
	\subfigure[]{\label{fig:0053_im}\includegraphics[width=0.23\textwidth, clip]{0053_im.eps}} \\ 
	\subfigure[]{\label{fig:0065_im}\includegraphics[width=0.23\textwidth, clip]{0065_im.eps}} \, 
	\subfigure[]{\label{fig:0070_im}\includegraphics[width=0.23\textwidth, clip]{0070_im.eps}} \, 
	\subfigure[]{\label{fig:0003_im}\includegraphics[width=0.23\textwidth, clip]{0003_im.eps}} \, 
	\subfigure[]{\label{fig:0055_im}\includegraphics[width=0.23\textwidth, clip]{0055_im.eps}} \\ 
	\subfigure[]{\label{fig:0013_im}\includegraphics[width=0.23\textwidth, clip]{0013_im.eps}} \, 
	\subfigure[]{\label{fig:0076_im}\includegraphics[width=0.23\textwidth, clip]{0076_im.eps}} \, 
	\subfigure[]{\label{fig:0059_im}\includegraphics[width=0.23\textwidth, clip]{0059_im.eps}} \, 
	\subfigure[]{\label{fig:0029_im}\includegraphics[width=0.23\textwidth, clip]{0029_im.eps}}    
	\figcaption{Normally-weighted images of the 16 detections. In order, 
	(a) 37W060, (b) 37W080, (c) 37W092, (d) 37W093, (e) 37W115, (f) 37W118, (g) 37W123, (h) 37W125, (i) 37W129, (j) 37W142, (k) 37W144, (l) 37W150, (m) 37W156, (n) 37W158, (o) B23, (p) B157.
	The first contours are at $\pm$3-sigma, increasing in increments of $\sqrt{2}$\label{fig:images}.}
\end{figure*}

After source identification, new calibrated visibility datasets were generated, centered on the position of the brightest pixel in the LR maps.
The package DIFMAP was then used to fit a circular Gaussian model to the visibilities.
The visibilities were finally re-shifted to the centroid of the Gaussian fit and final images were produced (\fig~\ref{fig:images}).

Flux measurements and their uncertainties were then derived using \blobcat\ \citep{Hales:2012} with the same parameters used by \citet{Middelberg:2013} (i.e. a pixellation error of only 1\% since the pixels are much smaller than the beamsize, and assuming a nominal surface brightness error of 10\%).
No errors associated with the primary beam correction are explicitly accounted for. 
37W060 is just beyond the radius probed by \citet{Middelberg:2013}, and given the primary beam correction factor of $>$ 23, it is possible that the flux error is greater than the nominal 10\%.

To characterise the source morphology, it was necessary to produce a robust estimate of the source size.
These visibilities were then averaged in frequency and polarization and averaged in time over 2 hours to produce approximately 200 visibilities per source.
These datasets were used to fit a Gaussian to the real part of the visibilities against $uv$ distance (corresponding to a circular Gaussian fit in the image plane).
The visibility weights and the scatter in the imaginary components were used to estimate the error bounds of each individual visibility.

The error on the fit (both flux and size) was then estimated using a perturbation analysis: a random amount of noise was added to the real component of each visibility (drawn from a Gaussian distribution with the standard deviation given by the error bound of that visibility) and the fit was recalculated.
This process was repeated 1000 times and the standard deviation on the distributions of the flux and the source width in the image plane provided the errors on these quantities.

\begin{figure*}[p]
	\centering
 	\subfigure[]{\label{fig:0099_model}\includegraphics[width=0.23\textwidth]{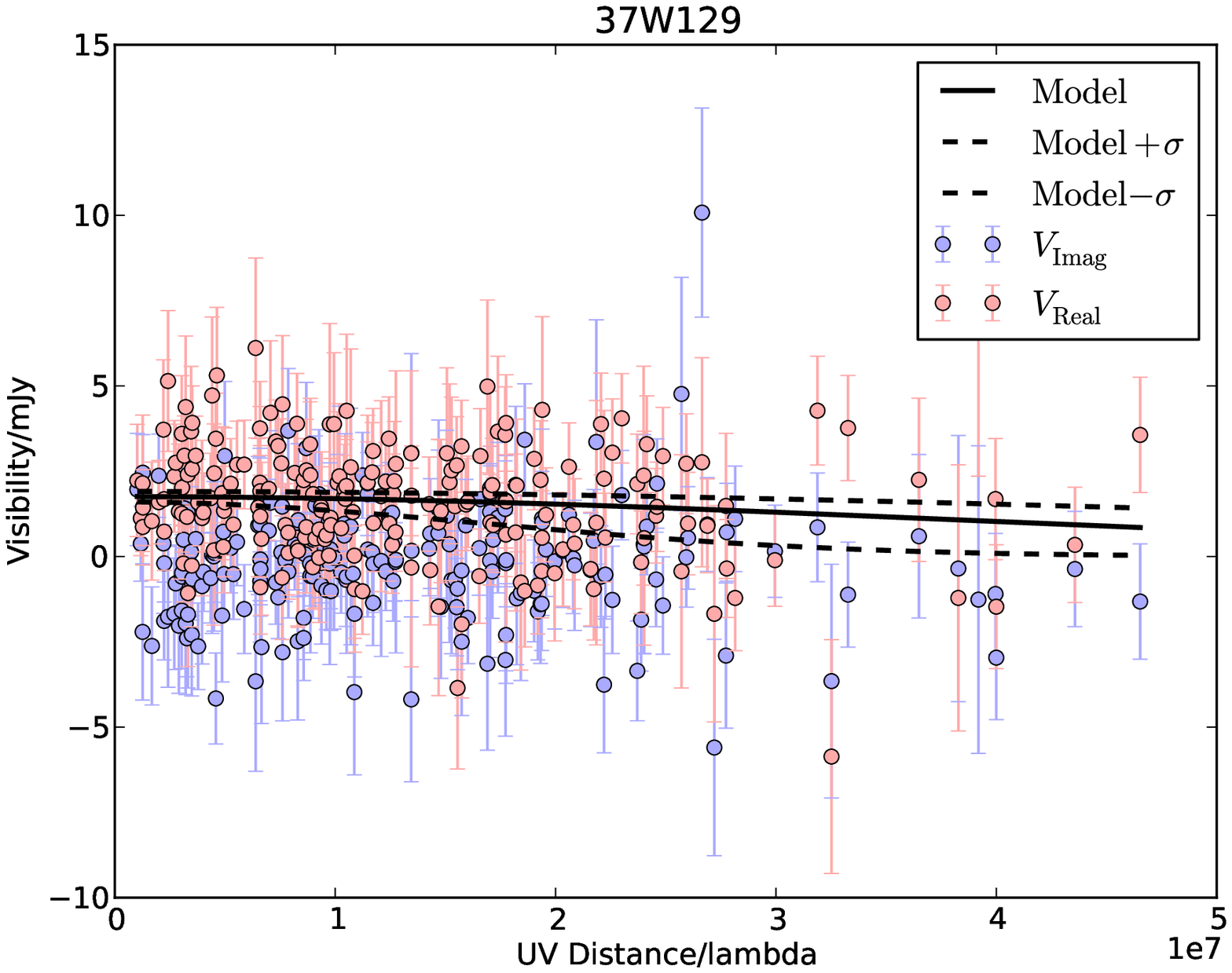}} \, 
 	\subfigure[]{\label{fig:0111_model}\includegraphics[width=0.23\textwidth]{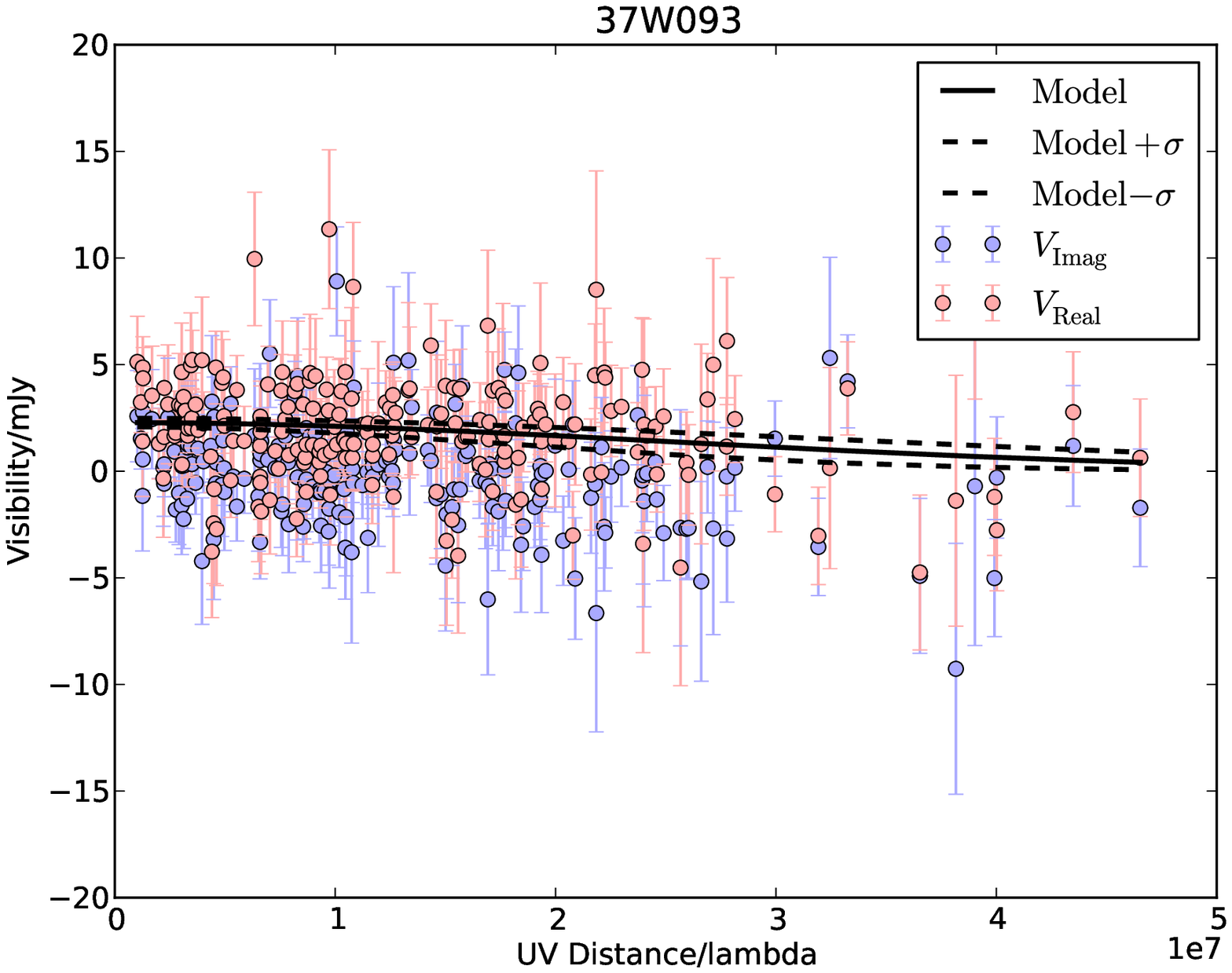}} \, 
 	\subfigure[]{\label{fig:0058_model}\includegraphics[width=0.23\textwidth]{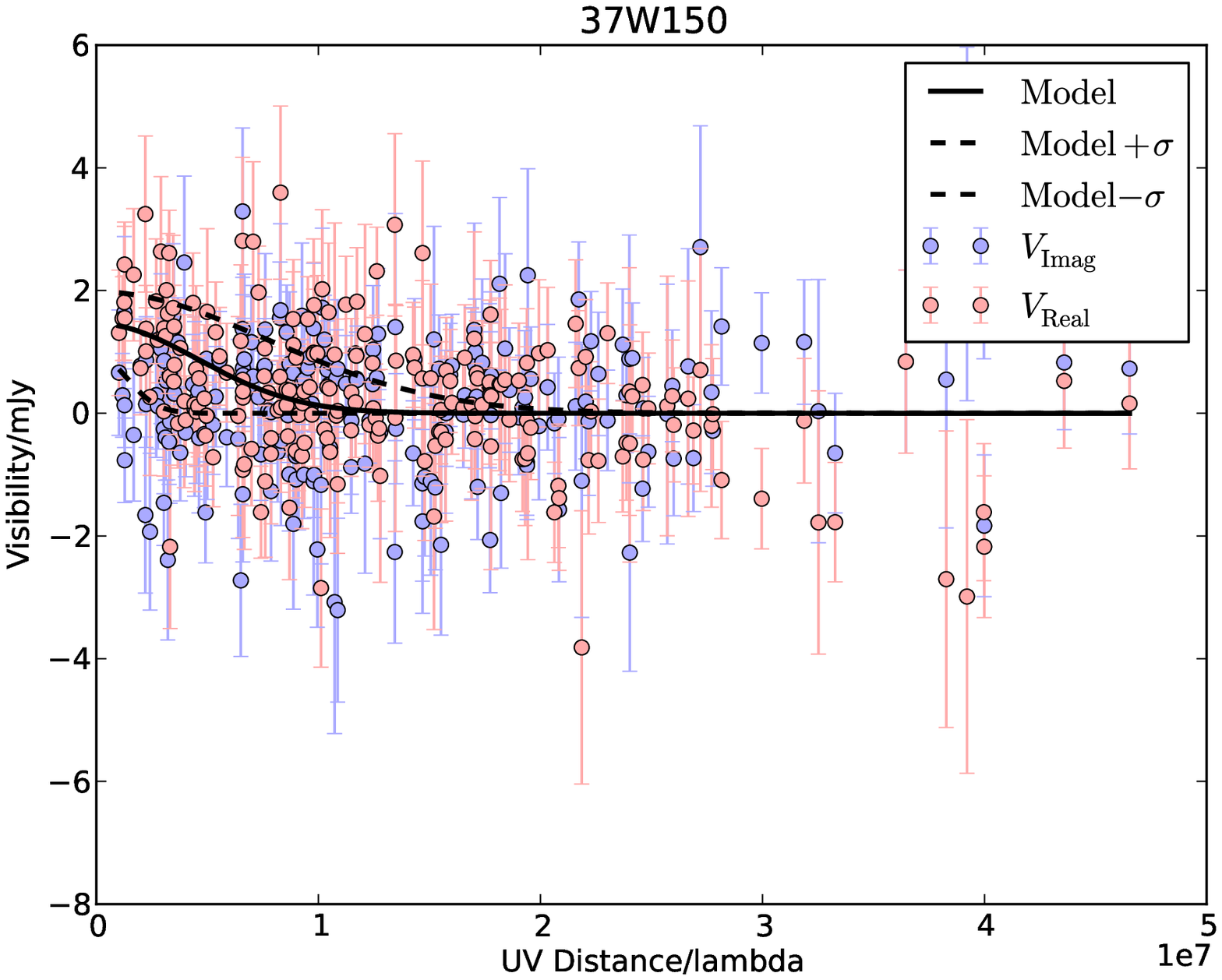}} \, 
 	\subfigure[]{\label{fig:0117_model}\includegraphics[width=0.23\textwidth]{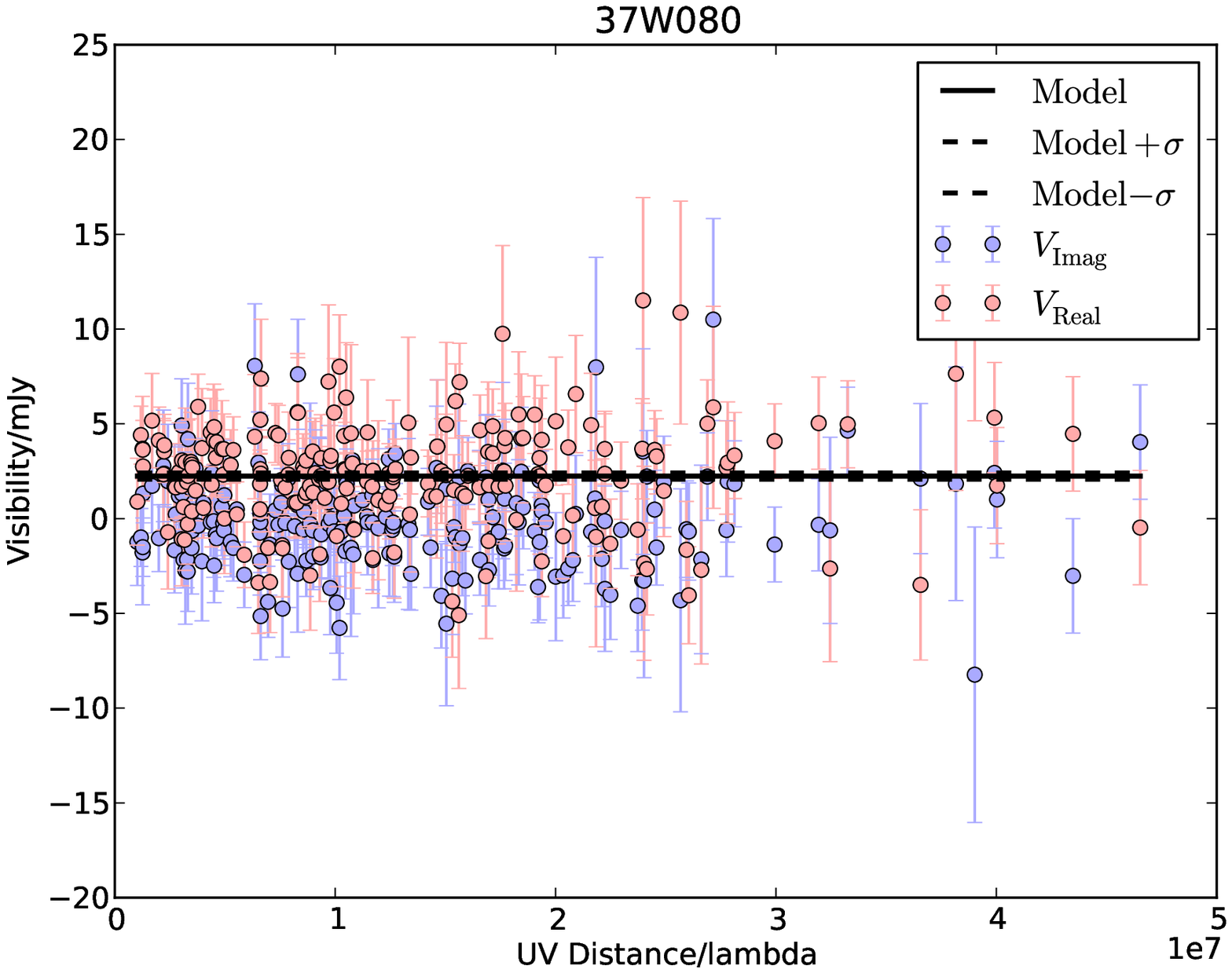}} \\ 
 	\subfigure[]{\label{fig:0154_model}\includegraphics[width=0.23\textwidth]{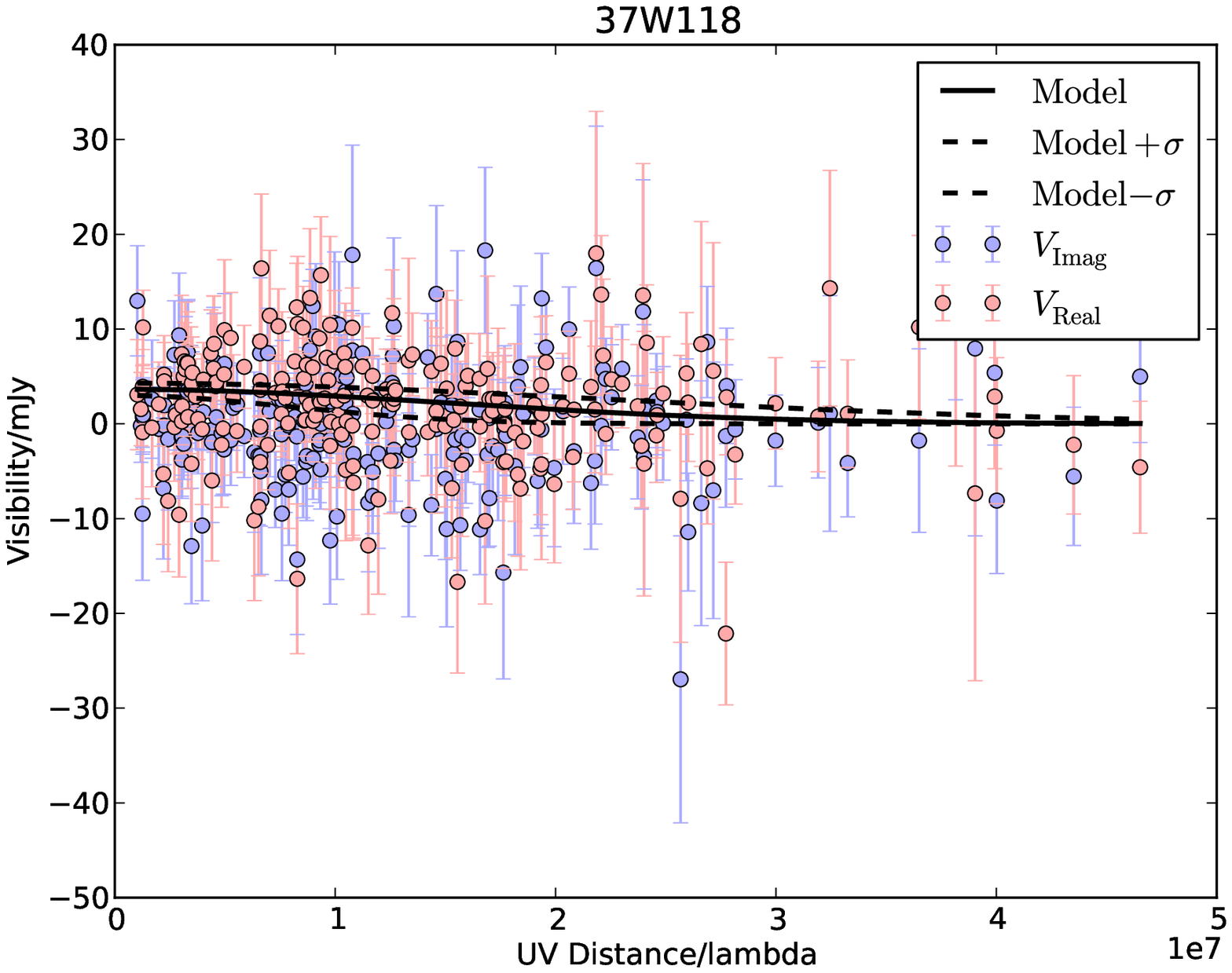}} \, 
 	\subfigure[]{\label{fig:0167_model}\includegraphics[width=0.23\textwidth]{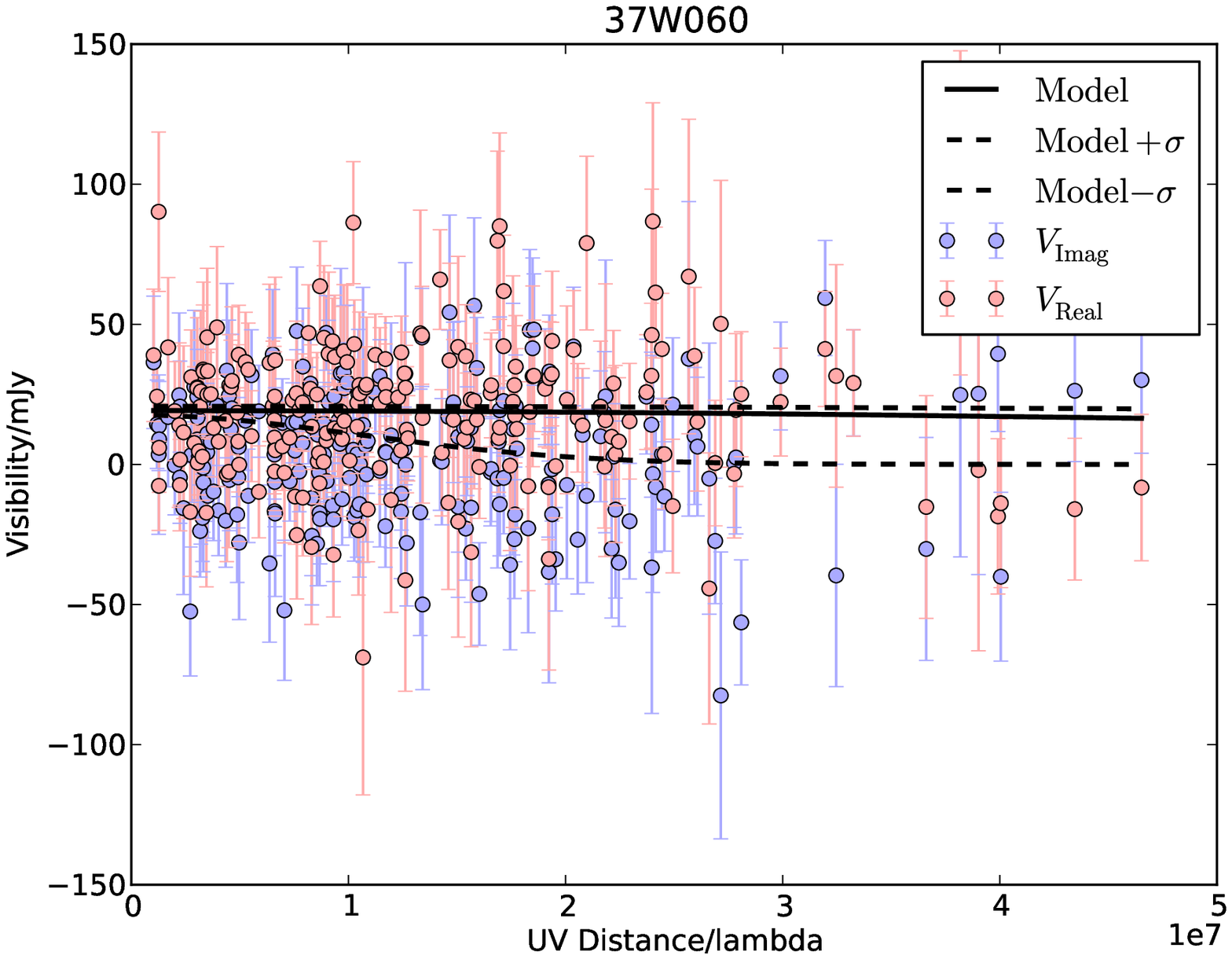}} \, 
 	\subfigure[]{\label{fig:0007_model}\includegraphics[width=0.23\textwidth]{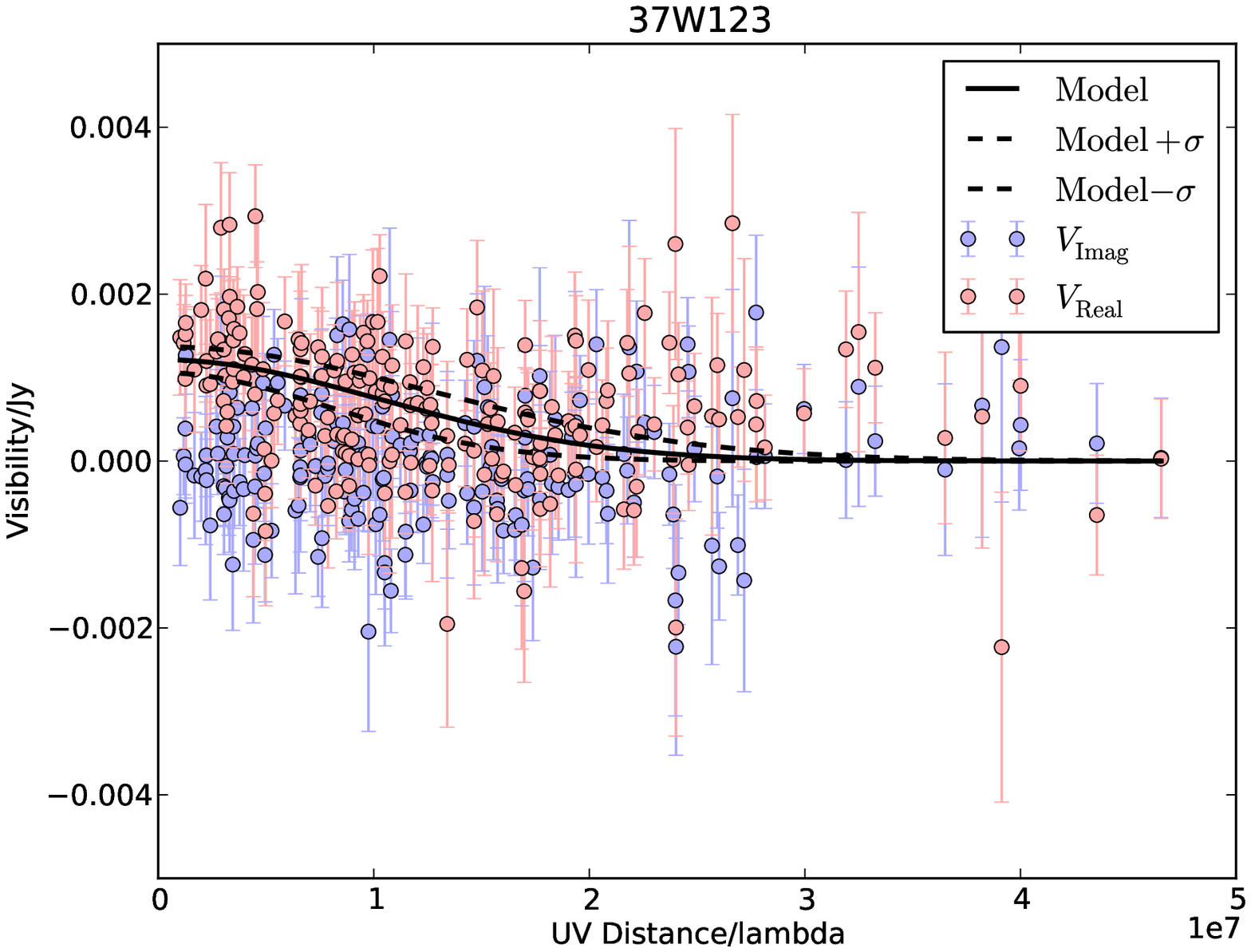}} \, 
 	\subfigure[]{\label{fig:0053_model}\includegraphics[width=0.23\textwidth]{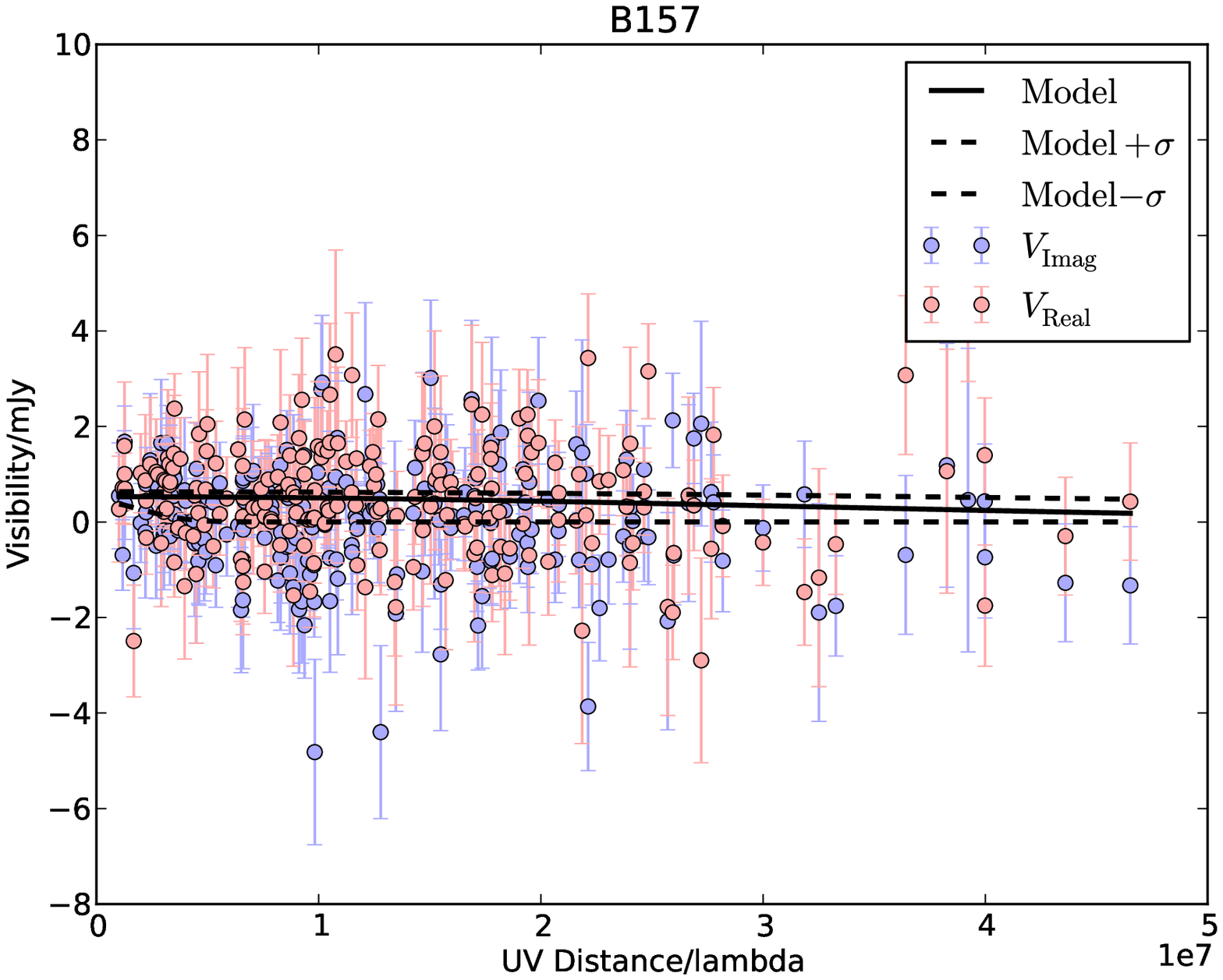}} \\ 
 	\subfigure[]{\label{fig:0065_model}\includegraphics[width=0.23\textwidth]{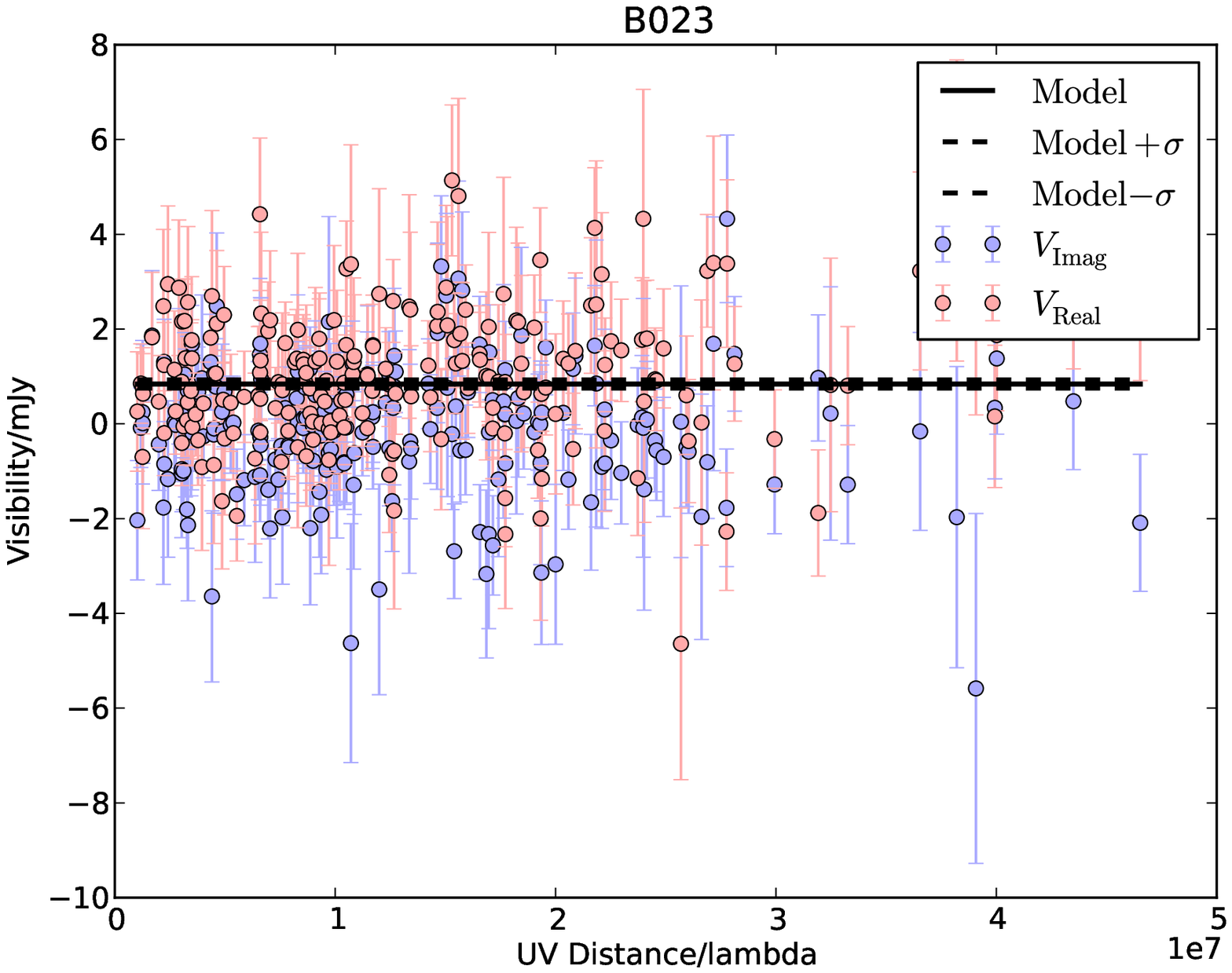}} \, 
 	\subfigure[]{\label{fig:0070_model}\includegraphics[width=0.23\textwidth]{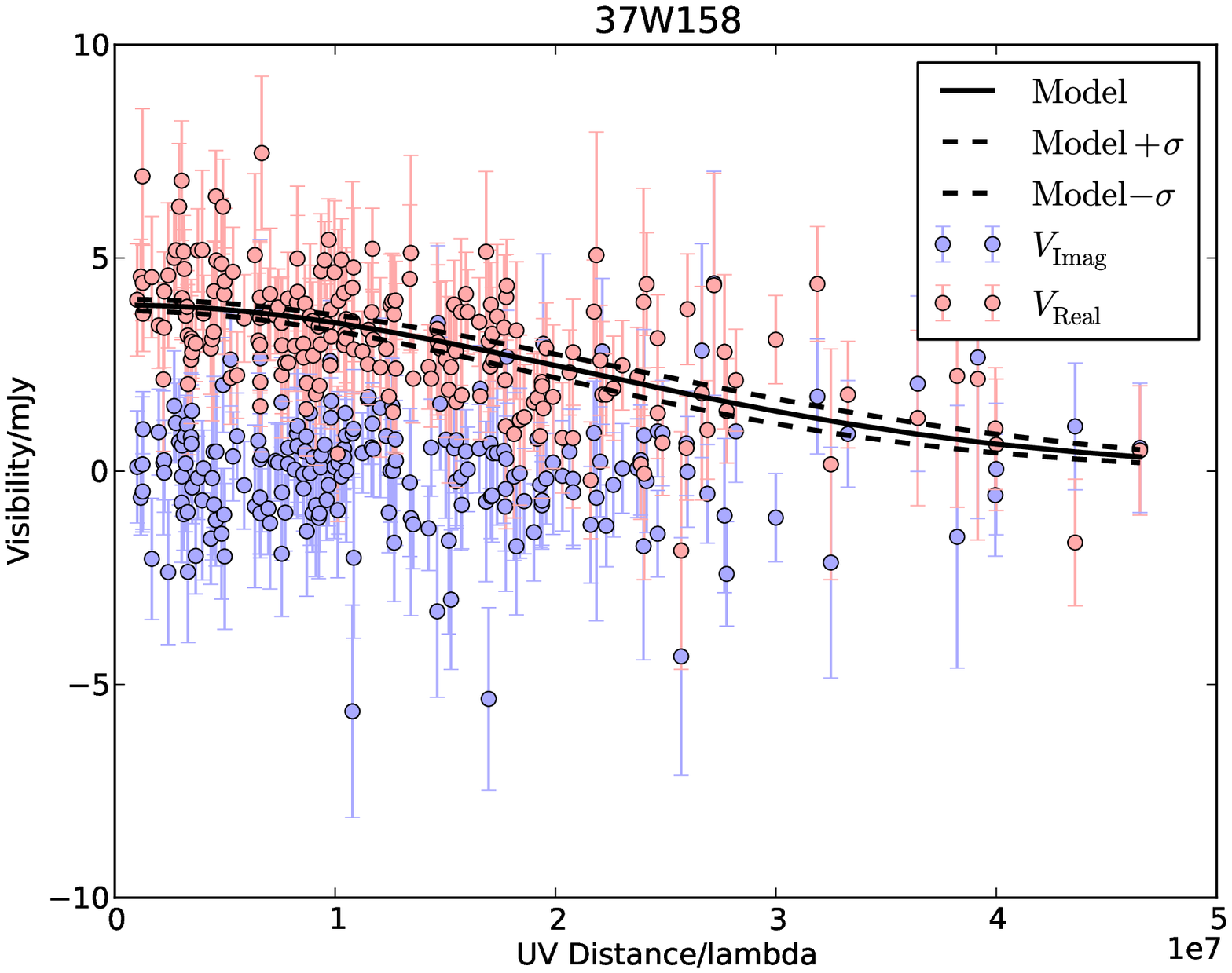}} \, 
 	\subfigure[]{\label{fig:0003_model}\includegraphics[width=0.23\textwidth]{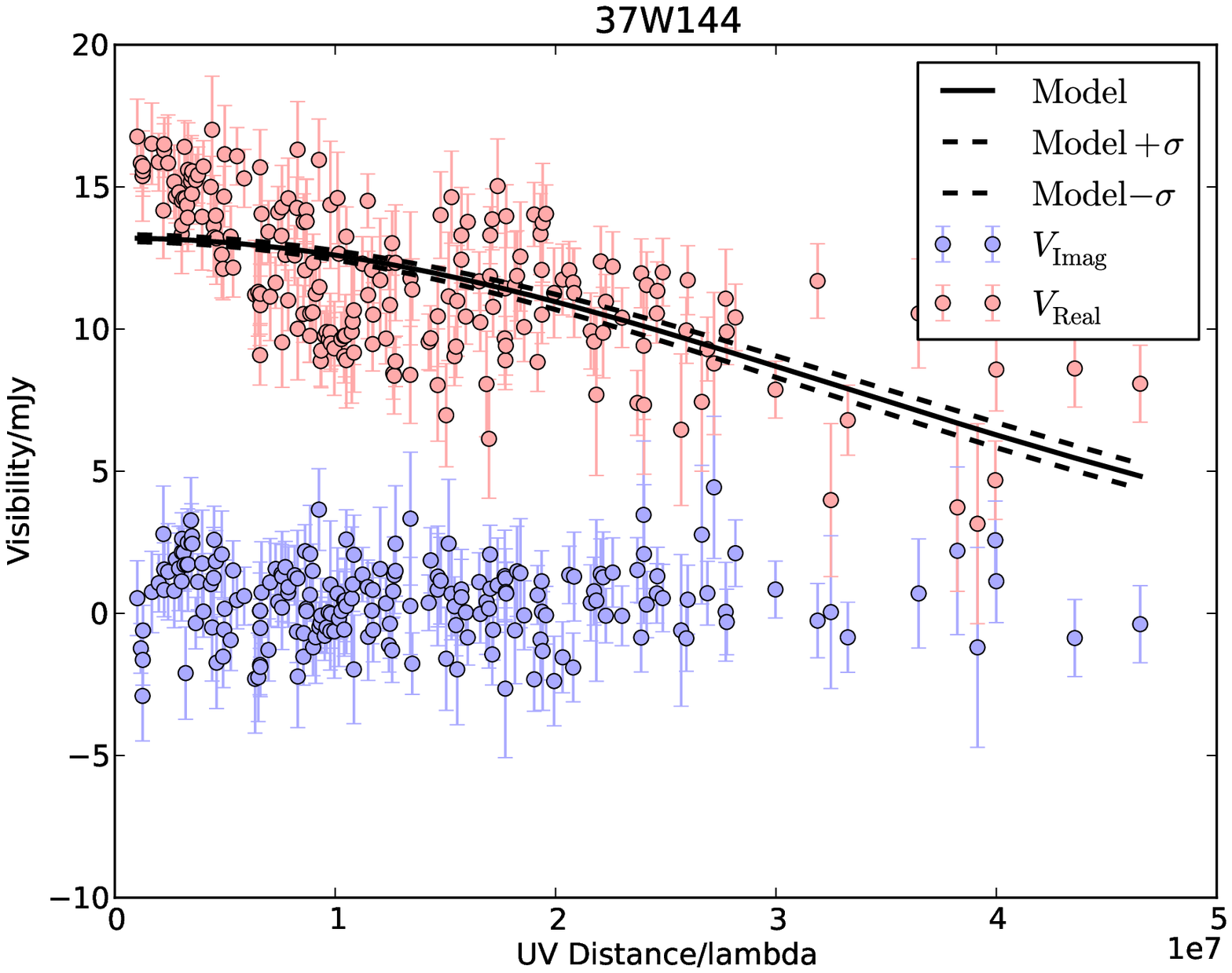}} \, 
 	\subfigure[]{\label{fig:0055_model}\includegraphics[width=0.23\textwidth]{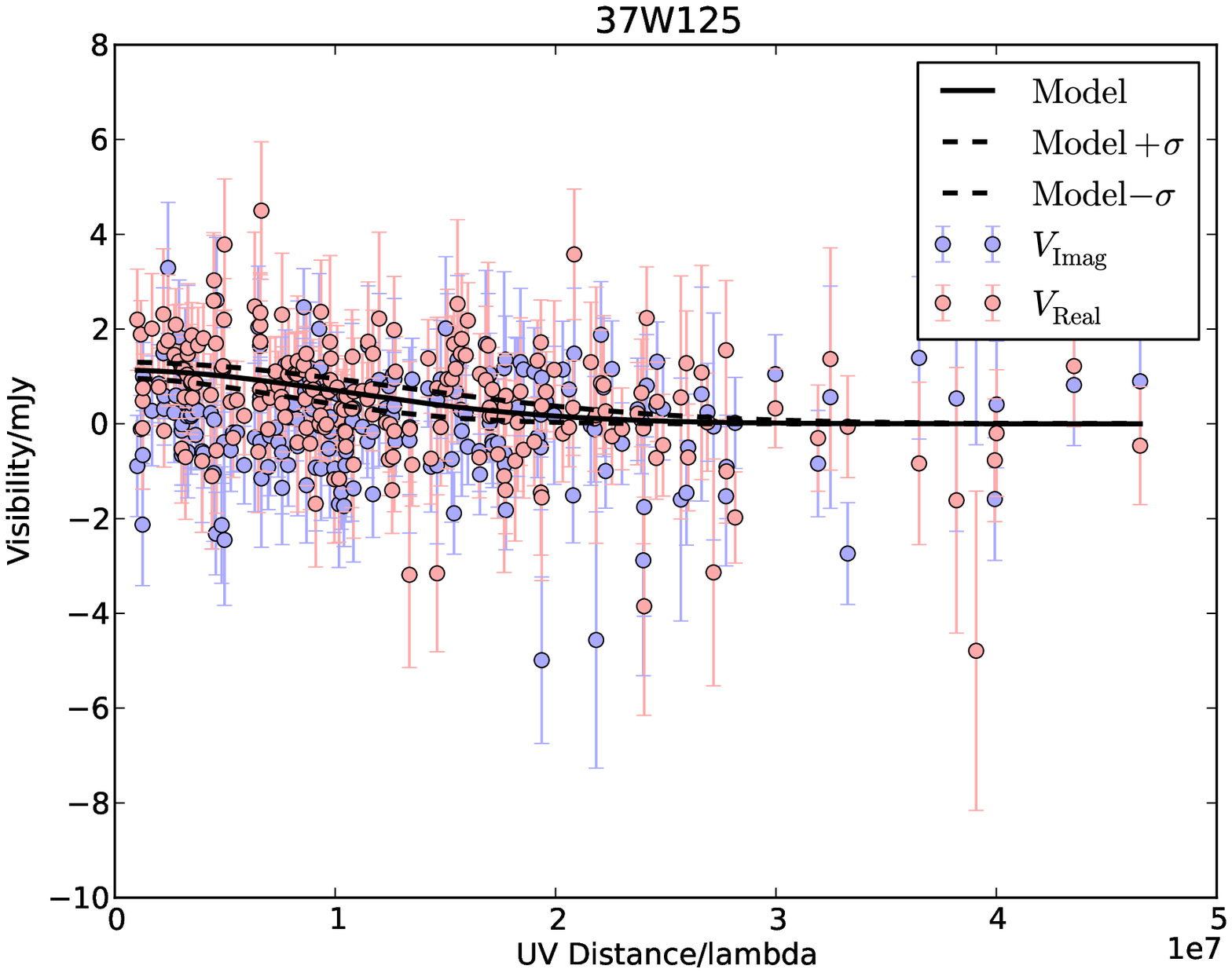}} \\ 
 	\subfigure[]{\label{fig:0013_model}\includegraphics[width=0.23\textwidth]{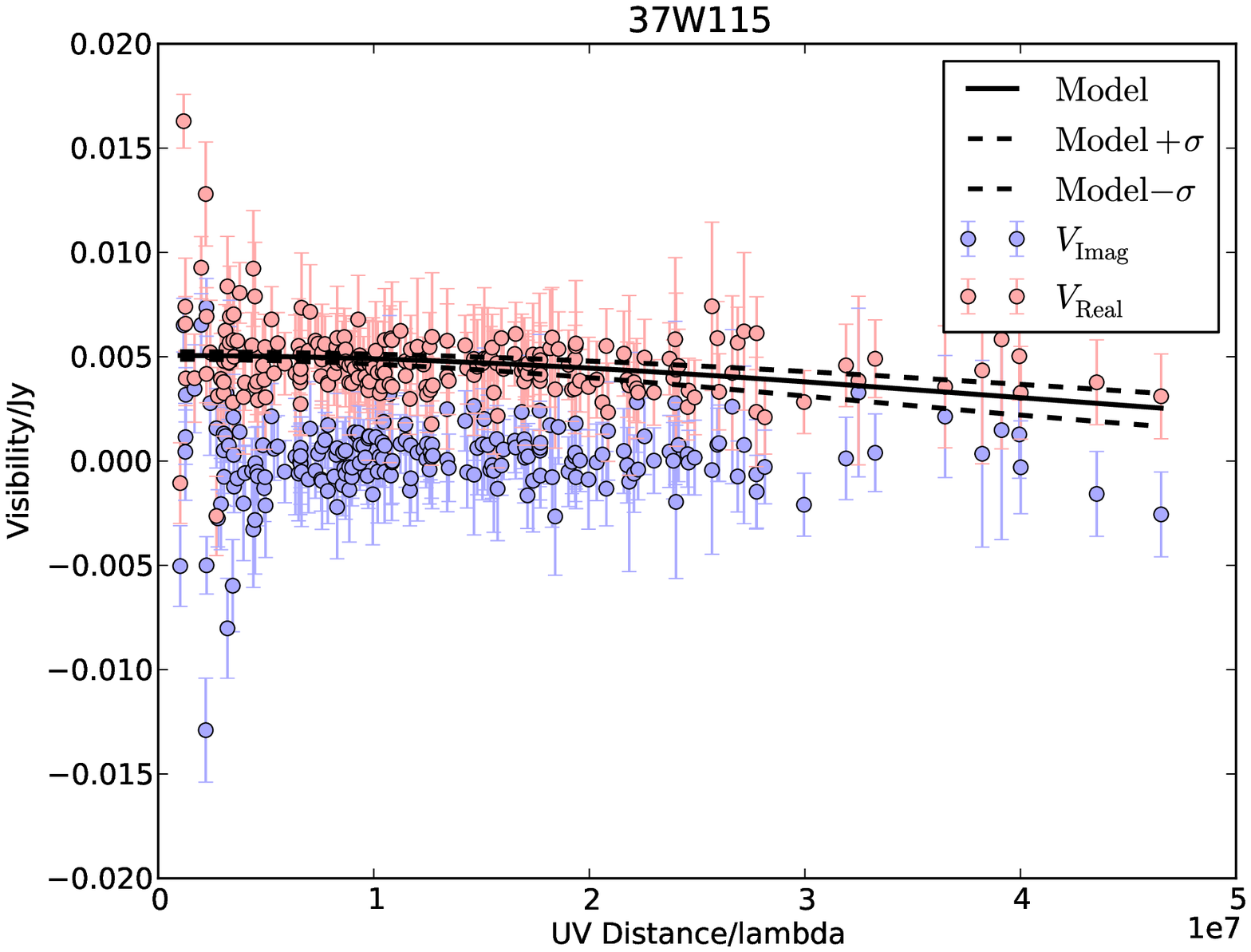}} \, 
 	\subfigure[]{\label{fig:0076_model}\includegraphics[width=0.23\textwidth]{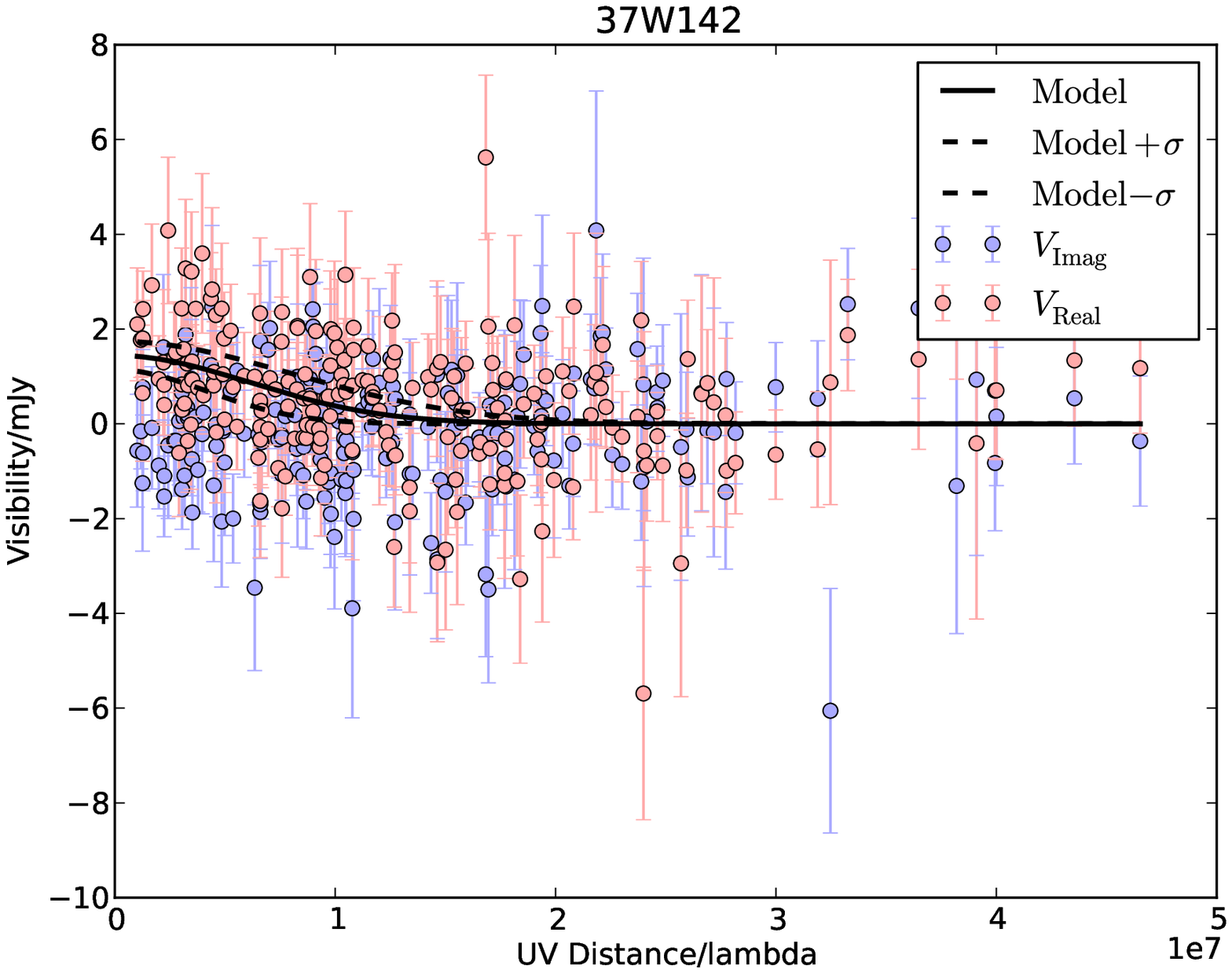}} \, 
 	\subfigure[]{\label{fig:0059_model}\includegraphics[width=0.23\textwidth]{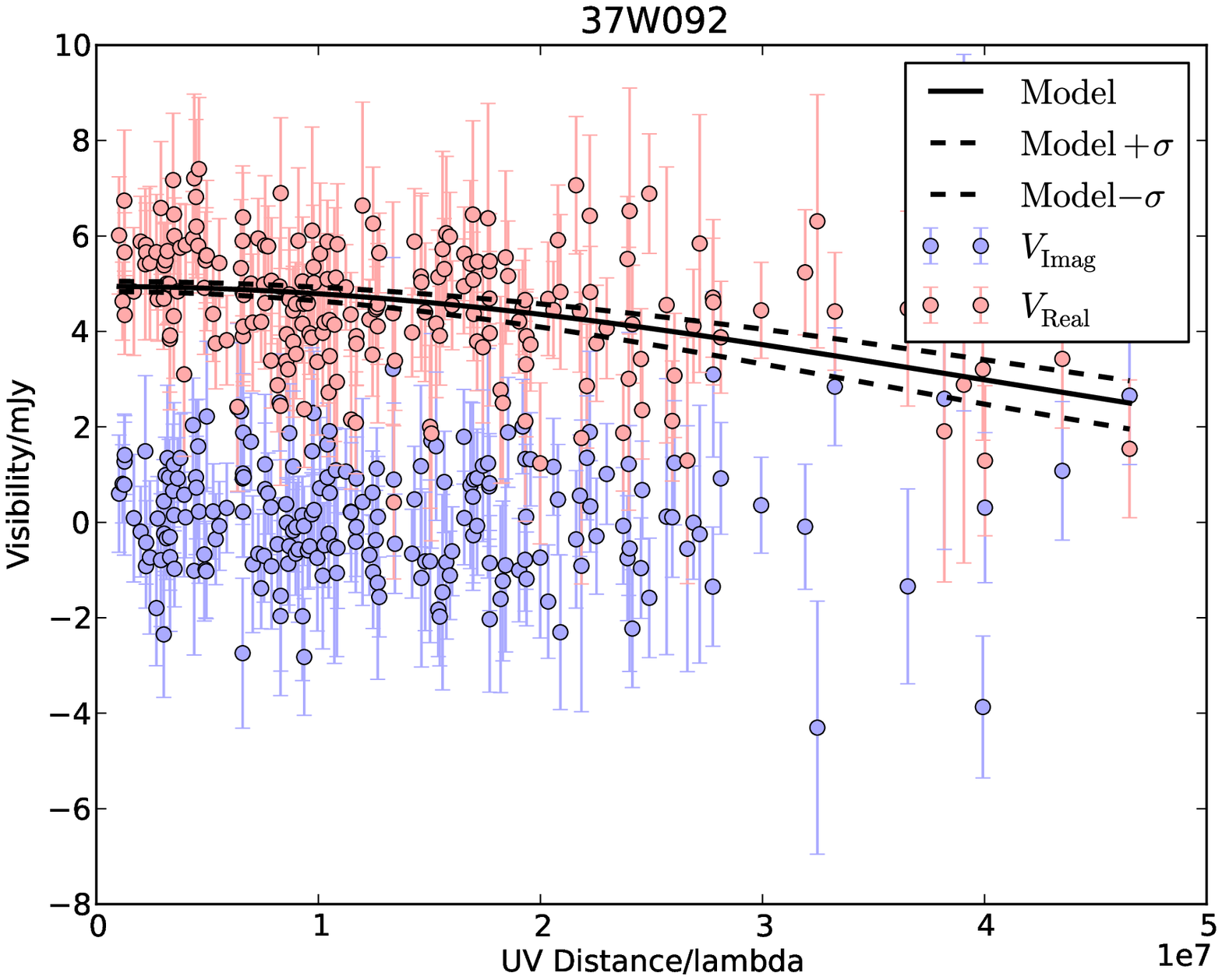}} \, 
 	\subfigure[]{\label{fig:0029_model}\includegraphics[width=0.23\textwidth]{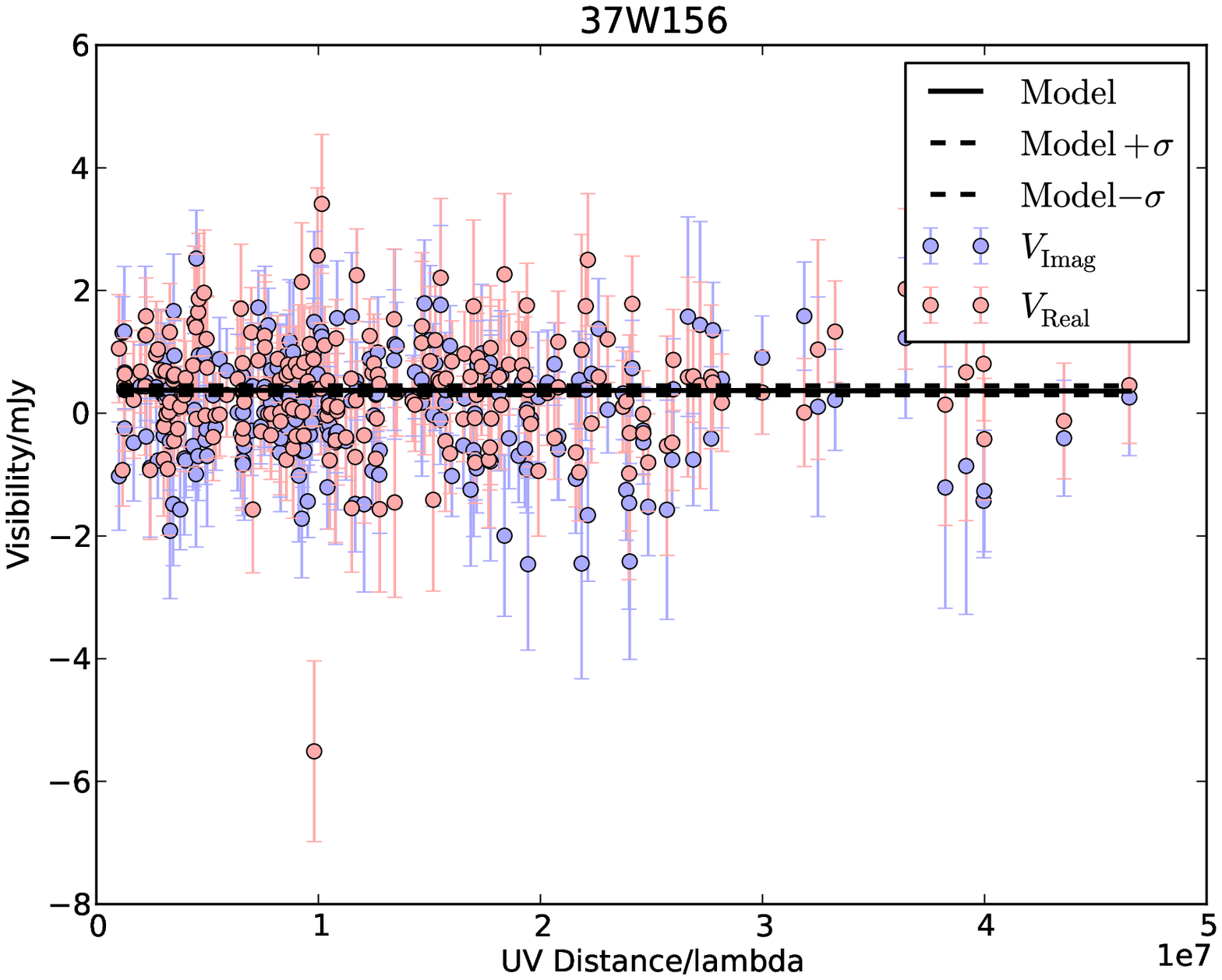}} \\ 
	\figcaption{Plots of the model fit of a Gaussian to the visibilities. In order, 
	(a) 37W060, (b) 37W080, (c) 37W092, (d) 37W093, (e) 37W115, (f) 37W118, (g) 37W123, (h) 37W125, (i) 37W129, (j) 37W142, (k) 37W144, (l) 37W150, (m) 37W156, (n) 37W158, (o) B23, (p) B157.
	The red dots show the real parts of the visibilities and their associated errors, the blue dots show the imaginary parts of the visibilities (with the same errors). The solid black line shows the least squares fit. The two dotted lines show this same fit with the standard error on both the zero-spacing flux and the Gaussian width first added, then subtracted.\label{fig:models}}
\end{figure*}

The final source positions and their errors are shown along with the \blobcat\ flux, fitted Gaussian size and zero-spacing flux in \tab~\ref{table:detections}.
The fit to the visibilities is shown in \fig~\ref{fig:models}.

\subsection{Source Astrometry}
The astrometric error on the source positions contains two components.
The first component is the formal error given by the limited S/N of the Gaussian fit.
The formal error for the position component of a Gaussian fit is given approximately by
\begin{equation}
	\label{eqn:poserr}
	\sigma_{\mathrm{pos}} = \delta_{\mathrm{pos}}\frac{\sigma_{\mathrm{rms}}}{2S}
\end{equation}
(see \citealp{Reid:1998}) where $\delta_{\mathrm{pos}}$ was the FWHM of the Gaussian fit, or the beamsize, whichever was larger\footnote{The beamsize for our observations was 9.77$\times$5.55\,mas with a PA 0.37\degr\, therefore 5.55\,mas was used as the minimum $\delta_{\mathrm{pos}}$ for the RA error, while 9.77\,mas was used for the declination error.}, $S$ was the zero-spacing flux from the Gaussian fit and $\sigma_{\mathrm{rms}}$ was the RMS noise in the source-detection map.

However, this formal error considers only the accuracy to which a Gaussian model can be fitted to the data and ignores any systematic shifts in the data itself.
For phase referenced VLBI observations at 18\,cm, there are two primary effects:
\begin{enumerate}
	\item The positional error of the primary calibrator source; and
	\item Large-scale structure in the ionosphere at the time of the observation, which leads to phase referencing errors, which displace the target sources.
\end{enumerate}

The primary phase reference source,  ICRF J003824.8+413706, has a catalog positional accuracy of 0.1 mas.
However, the ICRF positions are defined by observations at 2.3 and 8.4 GHz, and are known to be affected in some cases by source structure, both time invariant and time variable \citep[e.g.,][]{Porcas:2009}.
As shown in \fig~\ref{fig:calib}, ICRF J003824.8+413706 clearly has a core--jet morphology and is likely to be affected by this issue, especially since our observations are made at the lower frequency of 1.6 GHz.
Therefore, we consider it likely that the reference position used for a point--source model of ICRF J003824.8+413706 at 1.6 GHz is in error by 1--2 mas.

The magnitude of phase referencing errors is difficult to estimate, since they depend on the (difficult to model) ionospheric conditions at the time of the observation.
However, since the target--calibrator separation is quite small, only $\sim$1\degr, this contribution is likely to be relatively small.
Based on astrometric observations of pulsars at 1.6 GHz, a value of 0.25 - 0.5\,mas might be expected for a given observing epoch \citep[e.g.,][]{Brisken:2002}.
Thus, the total systematic error contribution is likely to be of order 1--2 mas, dominated by the uncertainty in the position and structure of the reference source ICRF J003824.8+413706.

For all sources, the formal position error estimate is smaller than this systematic error estimate, by a large margin in the case of brighter sources.
The use of the source 37W144 as an in--beam calibrator does not change these systematic error contributions, since the position derived for 37W144 was subject to these errors in the initial imaging.
In essence, these errors determine the offset of 37W144 from its true position, and the subsequent use of 37W144 as an in-beam calibrator then reinforces this offset.
However, when considered \emph{relative} to 37W144, the residual errors in positions for the other sources are much smaller.
Since the angular displacements are $<$20\arcmin, and there is no time interpolation of the solutions, the residual systematic errors are probably $<$0.1 mas, again drawing on experience from pulsar astrometry at 1.6 GHz \citep[e.g.,][]{Chatterjee:2009}.
For many sources, this is smaller than the formal fit uncertainty.

Accordingly, in \tab~\ref{table:detections} we list only the formal fit errors for positional quantities, and note that an additional uncertainty of 2 mas should be added in quadrature to each axis to obtain an absolute position error.
If future observations improve the accuracy of the 1.6 GHz position of ICRF J003824.8+413706 (and by extension 37W144), this error component could be reduced.

\subsection{Reduction of archival VLBA data}
The VLBA archive contains data from an 8-hour phase-referenced observation of 37W142.
The observation took place on 2003-09-19 and the same phase reference source was used as for our observations.
In addition 37W144 was observed intermittently as a check source.
Four 4\,MHz sub-bands were recorded, two at 18\,cm and two at 21\,cm.
Data were available for 9 antennas (Brewster being the exception).

Data reduction proceeded in much the same fashion as for our main observation with a couple of exceptions: before calibration, earth orientation parameters were updated with the correct values.
Other than the removal of all Owens Valley baselines, no flagging was deemed necessary and standard VLBA pulse cal corrections and system temperatures were used for phase reference calibration.
After fringe fitting, the 18\,cm and 21\,cm data were separated.
Several rounds of phase and amplitude and phase self calibration were carried out on the phase reference source and this was applied to the other sources.
The correlation position of the calibrator differed by approximately 0.3\,mas from that of the 2010 observations.
This position was assumed correct for this epoch.

37W144 was clearly detected and the steps described in \sect~\ref{sec:modeling} were carried out.
37W142 was marginally detected (which is consistent with our results) although the model fit failed to converge.

The results are shown in \tab~\ref{table:archive}.

\section{Results}
\label{sec:results}
\begin{deluxetable}{l l l c c c r r r r}
	\tablecolumns{10}
	\tablewidth{0pc}
	\tableheadfrac{0.05}
	\tabletypesize{\scriptsize}
	\centering
	\tablecaption{Detected Source properties and Cross-match with Other Surveys\label{table:detections}}
	\tablehead{
		\colhead{Name} & \colhead{RA (J2000.0)} & \colhead{Decl. (J2000.0)} & 
		\colhead{I$_{\mathrm{blob}}$} & \colhead{I$_{\mathrm{model}}$} & 
		\colhead{Model Size} &
		\colhead{Braun} & \colhead{37W} & \colhead{GLG} & \colhead{Stiele} \\
		\colhead{} & \colhead{} & \colhead{} & 
		\colhead{(mJy)} & \colhead{(mJy)} & 
		\colhead{(mas)} &
		\colhead{} & \colhead{} & \colhead{} & \colhead{} \\
		}
	\startdata
37W144  &00:42:51.6222130(12) &41:26:34.318347(23) &   20.6$\pm$1.0   & 13.20$\pm$0.12 &  2.67$\pm$0.11 &103    &144   &011   &1055  \\ 	
37W123  &00:42:25.919929(21)  &41:29:18.46241(42)  &  1.761$\pm$0.097 &  1.26$\pm$0.15 &  9.02$\pm$2.1  &69     &123   &\nodata &\nodata \\ 	
37W156  &00:43:18.635967(21)  &41:24:22.82067(42)  &  0.397$\pm$0.058 &  0.38$\pm$0.07 &  0.51$\pm$2.2  &131    &156   &\nodata &\nodata \\ 	
B157    &00:43:38.685540(22)  &41:30:26.56947(43)  &  0.649$\pm$0.073 &  0.53$\pm$0.10 &  2.75$\pm$2.6  &157    &\nodata &025   &1272  \\  	
37W125  &00:42:26.656261(16)  &41:18:04.94893(18)  &   1.70$\pm$0.11  &  1.13$\pm$0.17 &  8.63$\pm$2.1  &70     &125   &002   &\nodata \\  	
37W150  &00:43:04.037813(31)  &41:38:46.65434(35)  &   1.44$\pm$0.10  &  1.46$\pm$0.52 & 19.31$\pm$13   &118    &150   &\nodata &1123  \\  	
37W092  &00:41:45.1683974(28) &41:26:23.584359(56) &   7.63$\pm$0.39  &  4.94$\pm$0.11 &  2.20$\pm$0.29 &25     &092   &021   &706   \\  	
B23     &00:41:41.810643(19)  &41:29:36.45939(38)  &  0.896$\pm$0.086 &  0.84$\pm$0.08 &  0.00$\pm$0.40 &23     &\nodata &\nodata &\nodata \\  	
37W129  &00:42:32.252269(11)  &41:42:36.37584(21)  &   2.67$\pm$0.17  &  1.76$\pm$0.15 &  2.27$\pm$1.3  &71     &129   &046   &\nodata \\  	
37W142  &00:42:50.383045(25)  &41:16:10.13693(28)  &   2.77$\pm$0.16  &  1.44$\pm$0.30 & 14.35$\pm$4.1  &100    &142   &001   &\nodata \\  	
37W158  &00:43:26.3158109(39) &41:19:11.787621(77) &   5.09$\pm$0.27  &  3.90$\pm$0.13 &  4.17$\pm$0.36 &140    &158B  &005   &1224  \\  	
37W060  &00:40:35.778378(13)  &41:35:10.55738(26)  &   21.4$\pm$1.7   & 19.22$\pm$1.5  &  1.05$\pm$0.92 &\nodata  &060   &069   &\nodata \\  	
37W080  &00:41:18.052819(12)  &41:23:17.30580(25)  &   2.80$\pm$0.20  &  2.24$\pm$0.16 &  0.00$\pm$0.79 &4      &080   &\nodata &\nodata \\  	
37W093  &00:41:46.146853(11)  &41:16:39.24342(22)  &   3.15$\pm$0.20  &  2.28$\pm$0.20 &  3.47$\pm$1.0  &29     &093   &016   &715?  \\  	
37W115  &00:42:18.7006175(37) &41:29:26.297058(74) &   7.45$\pm$0.39  &  5.08$\pm$0.18 &  2.29$\pm$0.46 &61     &115   &019   &856   \\  	
37W118  &00:42:21.983656(21)  &41:08:00.99797(36)  &   5.37$\pm$0.43  &  3.66$\pm$0.64 &  5.85$\pm$2.8  &66     &118   &008   &\nodata \\  	
	\enddata
	\tablecomments{I$_{\mathrm{blob}}$ and I$_{\mathrm{model}}$  are the model and \blobcat\ flux as described in \sect~\ref{sec:modeling}.
The cross-match with Stiele 715 is only tentative (see \sect~\ref{sec:largescaleagn}).}
\end{deluxetable}
\begin{deluxetable}{l c c c c c c }
	\tablecolumns{7}
	\tablewidth{0pc}
	\tableheadfrac{0.05}
	\tabletypesize{\scriptsize}
	\tablecaption{Archival Data on 37W144\label{table:archive}}
	\tablehead{
		\colhead{Source} &
		\colhead{Wavelength}&
		\colhead{$\delta_{\mathrm {RA}}$} &
		\colhead{$\delta_{\mathrm{dec}}$} &
		\colhead{I$_{\mathrm{blob}}$} &
		\colhead{I$_{\mathrm{model}}$} & 
		\colhead{Model Size} \\
		\colhead{} &
		\colhead{(cm)} &
		\colhead{(mas)} &
		\colhead{(mas)} &
		\colhead{(mJy)} &
		\colhead{(mJy)} & 
		\colhead{(mas)} \\
		}
\startdata
37W144  	& 18	& 0.80$\pm$0.15	& -0.19$\pm$0.34	& 10.1$\pm$1.1	& 11.00$\pm$0.82	& 1.5$\pm$1.1	\\
37W144  	& 21	& 0.77$\pm$0.21	& 0.03$\pm$0.55		& 11.1$\pm$1.3	& 15.3$\pm$2.1		& 4.2$\pm$2.3	\\
\enddata
\tablecomments{$\delta_{\mathrm RA}$  and $\delta_{\mathrm Dec}$  give the difference of the fitted position offset from the positions derived from the sources in our own observations. The error is for the fit on the archival data only.}
\end{deluxetable}
\subsection{Cross-matching with other surveys}
Counterparts to the detected sources were searched for in the B90, 37W \citep{Walterbos:1985}, GLG \citep{Gelfand:2004} and Stiele \citep{Stiele:2011} catalogs.

\begin{figure*}[p]
	\centering
	\subfigure[]{\label{fig:b90err}\includegraphics[width=6.2cm]{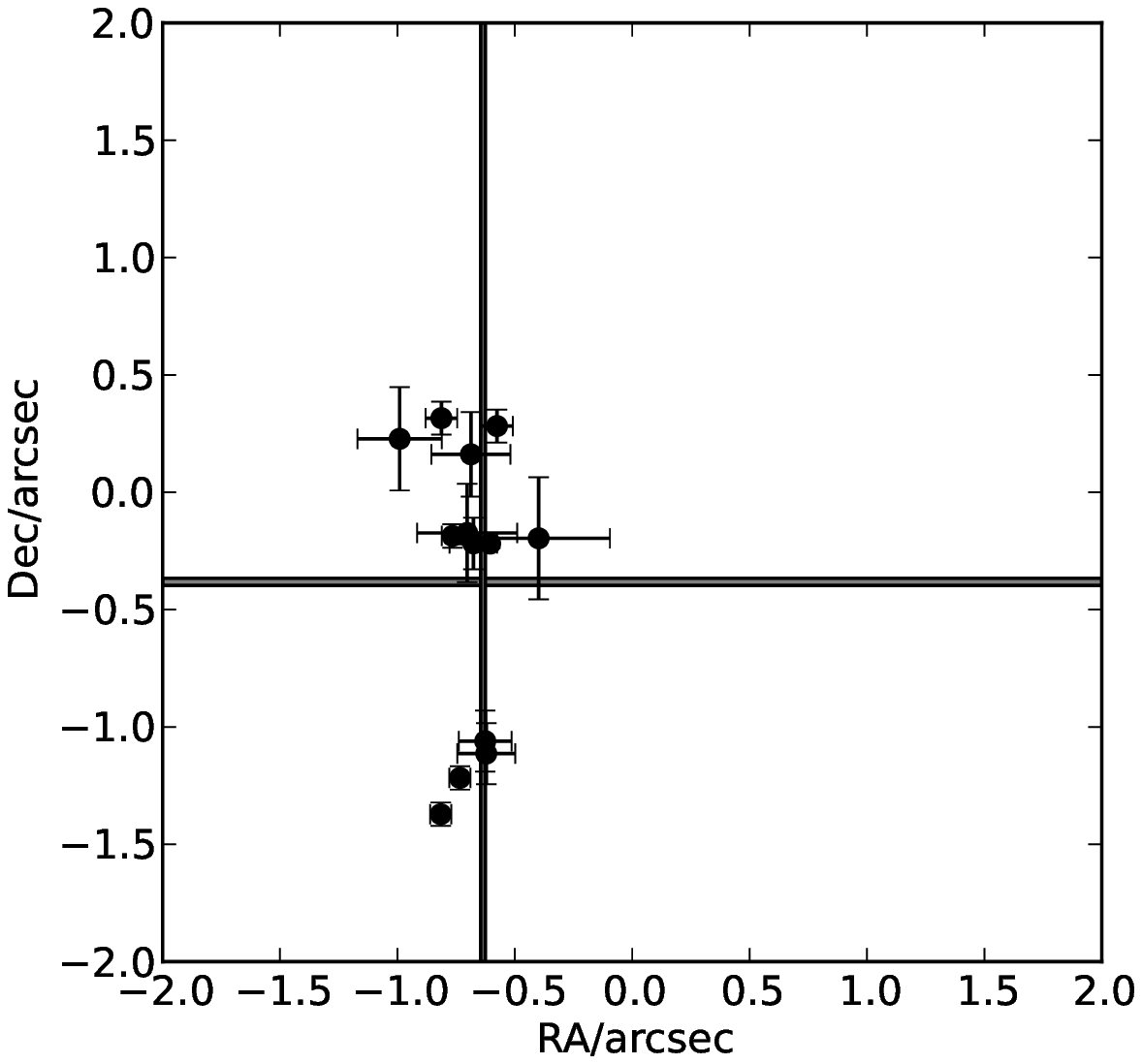}} \,
	\subfigure[]{\label{fig:37werr}\includegraphics[width=6.2cm]{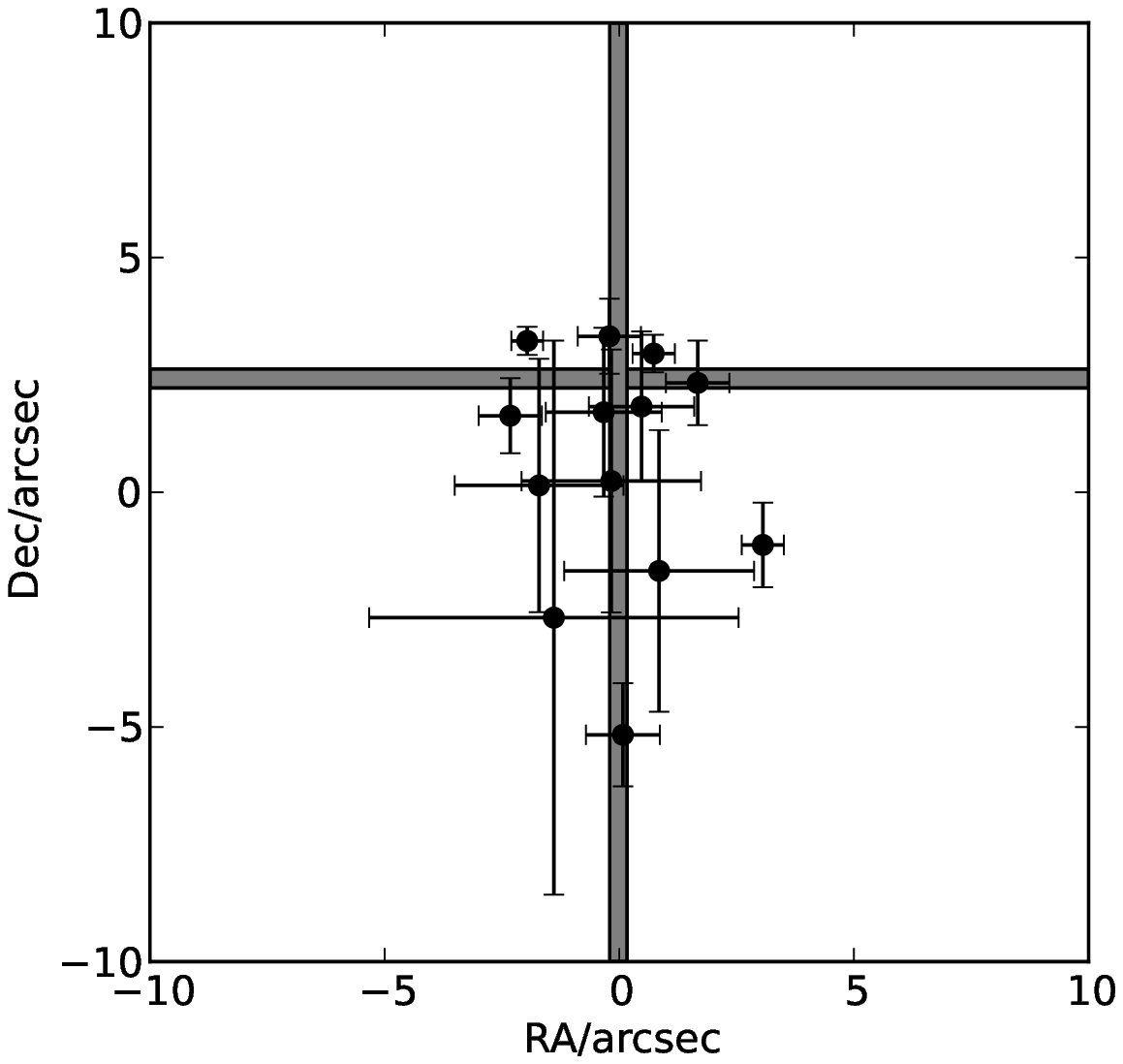}} \\
	\subfigure[]{\label{fig:glgerr}\includegraphics[width=6.2cm]{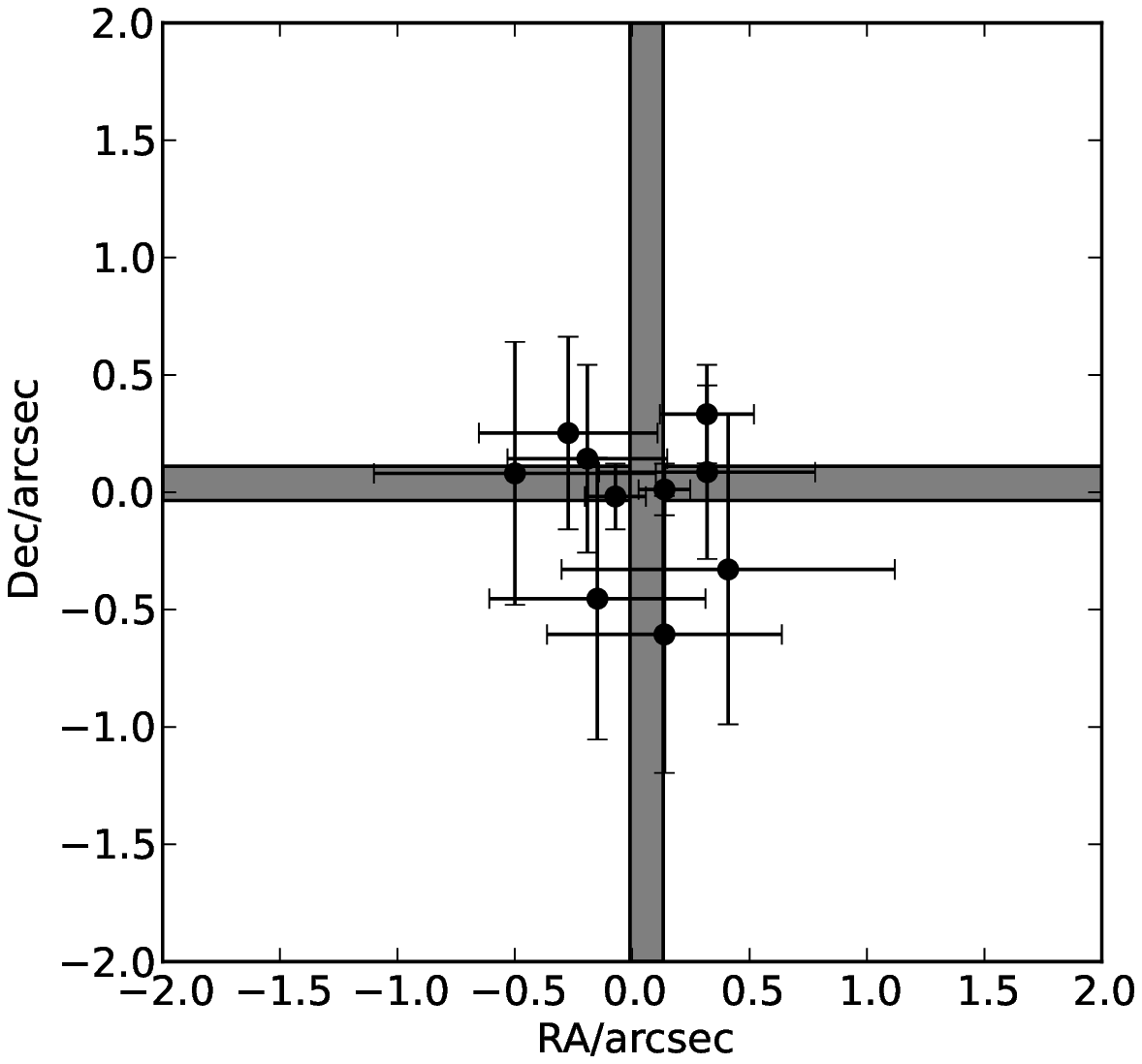}} \,
	\subfigure[]{\label{fig:stierr}\includegraphics[width=6.2cm]{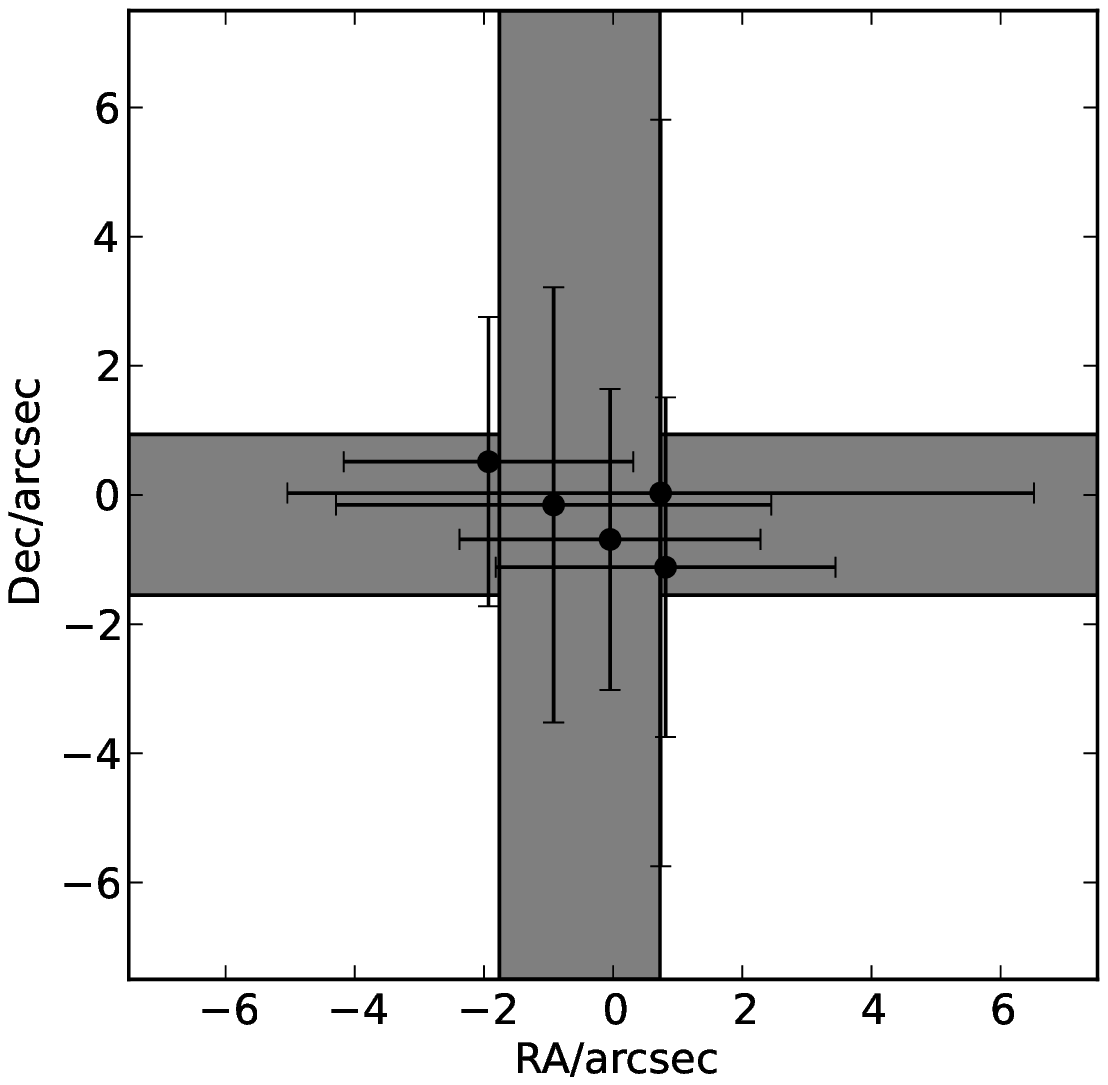}} \\
	\figcaption{Offset in position of sources in (a) Braun, (b) 37W, (c) GLG, (d) Stiele compared with VLBI positions. West is positive. The centroid of the gray lines is the weighted mean of the positional offsets. The width of the gray line is the cataloged errors for all sources combined in quadrature. 37W115 is excluded due to its arcsecond-scale structure\label{fig:survey_errors}}
\end{figure*}

The VLBI positions are compared with the positions of the counterparts in four surveys in \fig~\ref{fig:survey_errors}.
The B90 and the 37W catalogs appear to have systematic errors in the position in excess of the listed errors in RA and declination.
This is unsurprising given the considerable difficulties in achieving high astrometric accuracy in wide-field imaging (see, e.g. \citep{nvss:1998}, sect.~4.3).
However, they remain small relative to the minimum size we used for our source-finding analysis.

Using our source detection method with an input catalog with high astrometric accuracy, it would be possible to detect sources significantly below our detection threshold of 6.6$\sigma$.

\subsection{Source Classification}
Deciding whether the detected sources are intrinsic to M31 or background sources is vital if we are to draw any conclusions from our observations.
The only persistent background sources with brightness temperatures in excess of $10^6$\,K (and thus easily detectable by VLBI) are AGN.
Detecting 16 sources within one pointing of the VLBA is broadly consistent with \citet{Middelberg:2011}, so it would not be surprising if all of the detected sources were background AGN.
Indeed, all available data are consistent with this being the case.
In this section we consider the evidence for these sources being either intrinsic to M31, or background AGN.

All sources with X-ray counterparts have been classified by \citet{Stiele:2011} as AGN on the basis of their X-ray properties, on which basis we rule out the possibility that any of our detected sources are X-ray binaries or associated nebulae.

Young Supernova Remnants SNRs undergo free expansion for some time before beginning to decelerate.
Although there is much variation in measured expansion velocities, a typical expansion speed is 10\,000\,km\,s$^{-1}$.
This corresponds to $\sim$0.5\,mas year$^{-1}$ at the distance of M31.
Thus even the very youngest SNRs that are nonetheless old enough to be cataloged in B90 would likely be resolved into a ring or partial ring by the VLBA.
In addition, such recent supernovae would very likely have been observed.
Furthermore, SNRs older than $\sim$100 years would begin to be resolved out completely, though spots of compact structure within the ring could remain detectable on VLBI scales.

Pulsar Wind Nebulae (PWN) are another class of Galactic objects to which our observation would be sensitive.
As already discussed, the Crab nebula at the distance of M31 would have a total flux of approximately 6\,mJy and a total extent of 1\arcsec\, with filamentary structure on a scale of 15\,mas, as would a nebula associated with an X-ray binary.

However, none of our sources have a ring or partial ring morphology, and with one exception (37W115, which has AGN morphology), none show extended structure on scales greater than $\sim$10\,mas.
Moreover, with the exception of 37W115, the VLBI flux matches those from low-resolution surveys to within a factor of two, meaning that little or no extended structure is resolved out on scales to which the VLBA is insensitive ($\gtrsim 1$\arcsec).
While PWNe vary greatly in morphology and radio luminosity, this is smaller than the Crab nebula by two orders of magnitude.

While we cannot entirely rule out the possibility that one or more of our detections is a source intrinsic to M31, perhaps of a source population which does not exist in our own Galaxy, there is no real evidence for this.
Furthermore, many of the detected sources have features which are either suggestive of AGN, or which are incompatible with the galactic sources that we might expect to detect in M31.
These are summarized for each individual source below within four general categories:
background AGN on the basis of large-scale morphology;
background AGN on the basis of compact jet morphology;
background AGN on the basis of compact morphology;
background AGN on the basis of compactness and position.
\subsubsection{Background AGN on the basis of large-scale morphology}
\label{sec:largescaleagn}
37W158 and 37W093 appear to be the cores of double-lobed radio sources which are clearly visible in all low-resolution maps.
We note, however, that B90 and \citet{Walterbos:1985} treated the core and the lobes of 37W093 as separate sources, and that the source is located close to the starforming ring.
37W093 also seems likely to be associated with the X-ray source Stiele 715, though curiously it is offset by 18\arcsec\ from the VLBI position (a 2.8\,$\sigma$ error on the X-ray position).

\subsubsection{Background AGN on the basis of compact jet morphology}
Three of our sources (37W144, 37W123 and 37W115) show a clear core jet morphology in the VLBI maps and we classify them as background AGN on that basis.

\subsubsubsection{37W144}

37W144 has one of the highest flux densities of any of our sources and was used as an in-beam calibrator.
Though it appears to be only weakly resolved in our observations, we we do not recover all of the flux seen in the B90 catalog.
We also had access to VLA images at multiple frequencies and data from recent 8.4\,GHz VLBA observations of a microquasar within M31 \citep{Middleton:2013}.
The source is entirely unresolved in the VLA images and is almost flat spectrum (integrated fluxes are 33.2, 28.1, and 28.1\,mJy at 3.4, 5.3, and 7.5\,GHz respectively).

Self-calibrated 8.4\,GHz VLBA images revealed a bright core with a flux density of 10\,mJy beam$^{-1}$ with an extension to the South some 4\,mas in extent \citep[][Supp. Fig.~1]{Middleton:2013}.
A small amount of additional data on this source was available in the VLBA archive (see \tab~\ref{table:archive}).
The derived position and morphology of the source are entirely consistent with our observations, however the flux is significantly lower.

\citet{Gelfand:2005} suggest this source as a Pulsar Wind Nebula (PWN) candidate on the basis that it is within the optical disk and has a supernova remnant (SNR) classification (though the candidate SNR optical counterpart is almost 1\arcmin\ away).
The fact that most of the B90 flux is concentrated within 4\,mas (=0.014\,pc) strongly argues against this being a Pulsar Wind Nebula.

\citeauthor{Gelfand:2005} also classify this source as a High-frequency variable, since it varies by $(21\pm2)$\% between the B90 and NVSS \citep{nvss:1998} surveys.
Its 325\,MHz flux also varies by $(51\pm15)$\% between the GLG and WENSS surveys.
We see further evidence for this variability in that the current VLA 3.4\,GHz flux is 30\% higher than the Braun flux.
There is also a difference in brightness of a factor of two between our VLBA data and that from the VLBA archive.
This level of variability is consistent with the source being a background AGN.

\subsubsubsection{37W123}
37W123 shows a core-jet structure with the jet some 8\,mas east and slightly north of the core.
\subsubsubsection{37W115}

\begin{figure}[t]
	\centering
	\includegraphics[clip, width=80 mm]{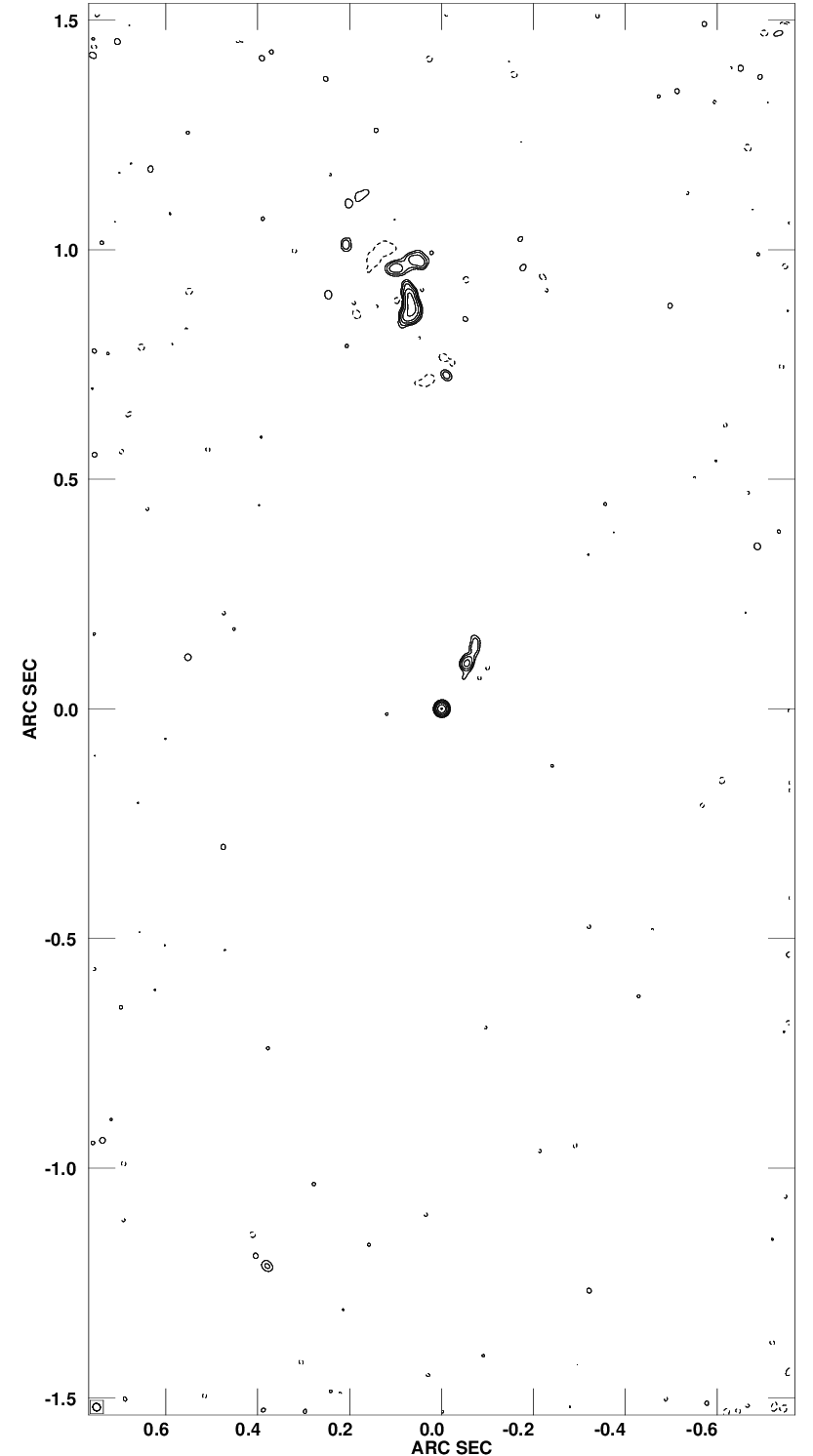}
	\figcaption{\label{fig:37W115} Image of 37W115 with a $uv$ cutoff of 10M$\lambda$ to show large-scale structure.}
\end{figure}

This source is unresolved in B90, however 2\,cm VLA maps \citep[0039+412 in ][as one of the B3-VLA CSS sample]{Rossetti:2006} reveal a core with evidence of a jet extension and two lobes.
All of these features (core 5.22$\pm$0.52\,mJy, jet 3.83$\pm$0.38\,mJy, northern lobe 23$\pm$2.3\,mJy, and southern lobe 0.84$\pm$0.08\,mJy) are visible in our map (\fig~\ref{fig:37W115}), however we recover only 12\% of the B90 flux.
It is very likely that most of this missing flux in the lobes which are largely resolved out by the VLBA, leaving only small hot-spots visible.

\citeauthor{Gelfand:2005} classify this source as a ``high-frequency variable'' though it only varies by $(6\pm4)$\%.
Since the lobes would not be expected to be variable, this would represent an extremely high level of variability were it attributable only to the core.
It is more likely to be due to experimental error.

\subsubsection{Background AGN on the basis of compact morphology}
37W142 and 37W125 are both located close to the core of M31 (1.1\arcmin\ and 4.8\arcmin\ from M31* respectively), while 37W150 is located on the edge of one of the largest H\II\ regions in M31.
In addition, the diameter of the circular Gaussian fit of these three sources is significantly larger than those of the other sources (though the fit for 37W150 is extremely poor, as evidenced by the large error).

\subsubsubsection{37W142}

The visibilities for both our data and the archival data fit a circular Gaussian.
In the images there is some suggestion of extended structure, however none exceeds 4\,$\sigma$ and it is nonetheless confined within some 30\,mas.

The VLBA flux matches the B90 flux, thus if this were a hot-spot on the edge of a SNR, there is no further radio emission.
The lack of any X-ray counterpart also argues against this hypothesis.

\subsubsubsection{37W125}
37W125 is classified by \citeauthor{Gelfand:2005} as both a ``steep spectrum'' source ($\alpha=-0.45\pm0.34$ and a ``normal galaxy'' based on its 2MASS counterpart and moderate spectral curvature ($\varphi=0.13\pm0.05$).
In our data it is well-described by a circular Gaussian 10\,mas in extent.
The inferred spectral index and curvature are small compared to their errors and could be due to source variability.

\subsubsubsection{37W150}
%

This source lies just on the edge of one of the larger and brighter H\II\ regions in the ring.
It lies extremely close to two H\II\ regions cataloged by \citet{Walterbos:1992}.
Although the source is only weakly detected, there is evidence of some extended structure, possibly jet-like.

It has an X-ray counterpart which is classified as an AGN based on its X-ray spectrum.

\subsubsection{Background AGN on the basis of compactness and position}
The remaining sources (37W156,  B157, 37W092, B23, 37W129, 37W080, 37W118, 37W060) are unresolved and are located well away from the core, starforming ring and significant H\II\ regions of M31.
Several of them show some evidence of jet morphology.
These are all likely to be background AGN.

5 of these sources have fluxes (as measured by \blobcat) significantly ($>1.5\sigma$) brighter than that in B90.
Two of these five sources have previously shown evidence of variability.
37W118 has a 37W flux which is 2.7$\sigma$ higher than the B90 flux but consistent with our own.
Likewise, 37W129 has an NVSS flux which is 1.7$\sigma$ higher than the B90 flux but consistent with our own.

This is a much higher fraction of sources with VLBI flux $>$ VLA flux than that found by \citet[][see \fig~14]{Middelberg:2013}.
This may be due to the two-decade gap between the B90 observations and our own.
An alternate possibility is that the error on the absolute flux density scale in one or the other catalog is underestimated.

\subsection{Stability of ICRF J003824.8+413706}
\begin{figure*}[t]
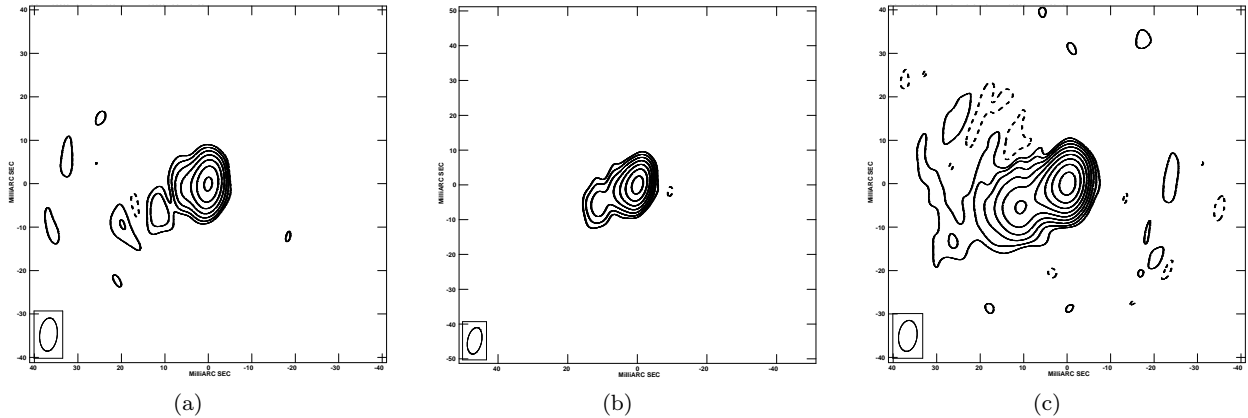

	\centering
	\resizebox{17cm}{!}
	{
		\subfigure[]{ 
			\label{fig:calibbk105}
			\resizebox{5cm}{!}{\includegraphics[clip]{J0038-18CM_4_im.eps}}}\quad 
		\subfigure[]{
			\label{fig:calibba097_match}
			\resizebox{5cm}{!}{\includegraphics[clip]{J0038_MATCH.eps}}}\quad 
		\subfigure[]{
			\label{fig:calibba097}
			\resizebox{5cm}{!}{\includegraphics[clip]{J0038+41_8_im.eps}}} 
	}
	\figcaption{A comparison of the phase reference source ICRF J003824.8+413706  at 18\,cm for (a) VLBA archival data (Epoch 2003-September-19) with contours starting at $\pm3\times$ the rms value of 0.47\,mJy; (b) our data (Epoch 2010-07-4) with contours chosen to match the former; (c) our data with contours starting at $\pm3\times$ the rms value of 0.089\,mJy. Contours progress in factors of 2.\label{fig:calib}}
\end{figure*}

\Fig~\ref{fig:calib} shows images of the phase reference calibrator in archival data compared with our data.
A modest increase in flux is seen between epochs (301$\pm$30 mJy vs. 386$\pm$39 mJy) and there is some evidence for evolution of source structure.

This is reflected in the change in the ICRF position between the epochs of approximately 0.3mas SE.
In contrast, we have several sources which show no extended structure whose current measurement errors are significantly smaller than this.
The use of these sources, or an ensemble of them is likely to produce a much more robust reference frame for multi-epoch astrometric observations of M31 than the ICRF source alone.

\subsection{Possible scatter-broadening by the ISM of M31}
\begin{figure*}[t]
	\centering
	\includegraphics[clip, width=\textwidth]{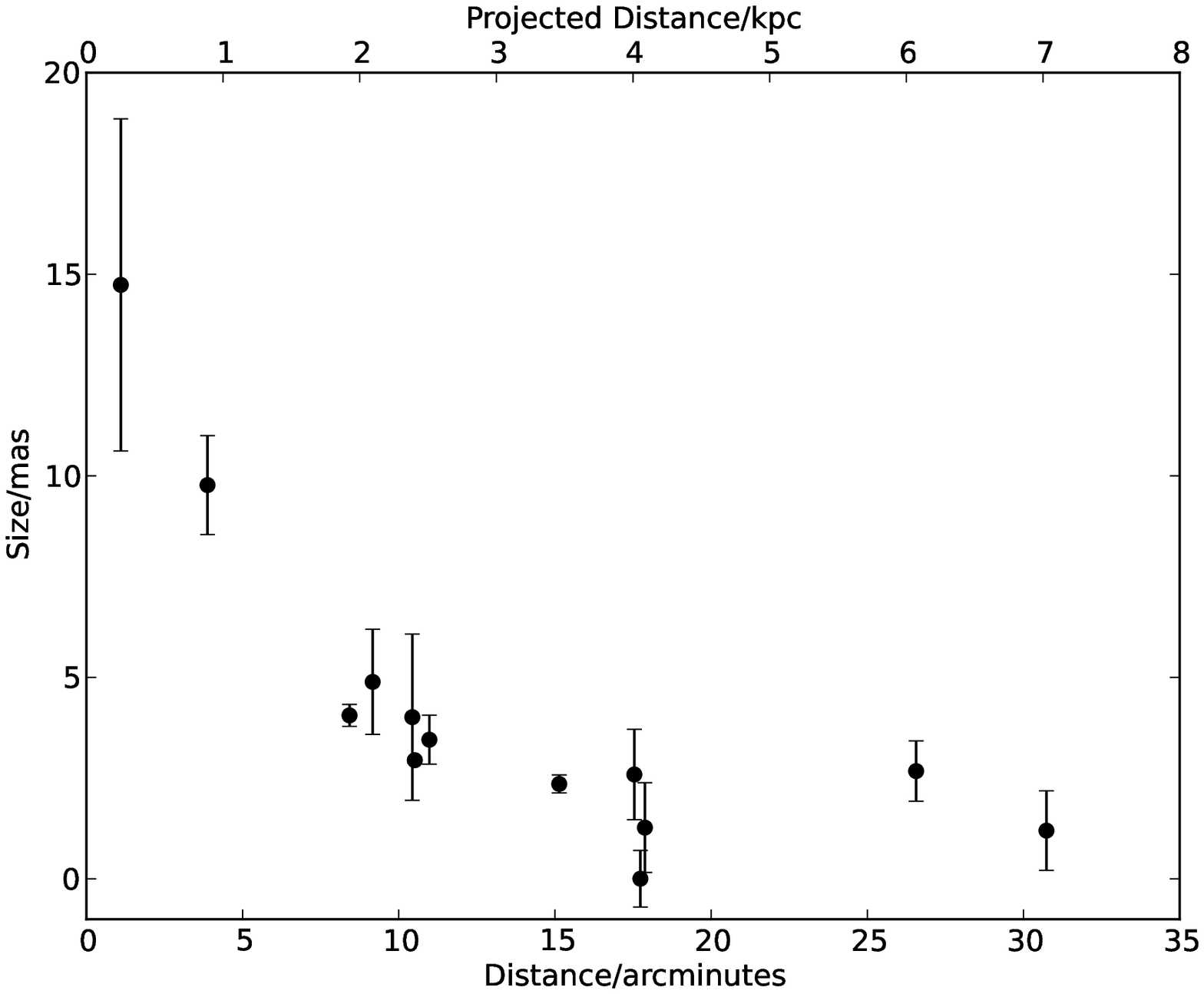}
	\figcaption{Size of sources as determined by circular Gaussian fit plotted against distance from M31*. 37W123, 37W115 and 37W150 are excluded from this plot due to their extended structure (37W150 is also excluded due to the large error on its size).\label{fig:scattering}}
\end{figure*}

\Fig~\ref{fig:scattering} shows the size of each source as determined by model fitting of the visibilities, plotted against its projected distance from M31* (37W123 and 37W115 are excluded from this analysis on account of their mas-scale extended structure.
37W150 is excluded on the basis of the poor constraint of its fit).
It is immediately obvious that the two sources closest to M31* (37W142 and 37W125) have the largest size.

Images of 37W142 and 37W125 show no evidence of complex structure (such as the jet of 37W123).
On the contrary, there is no evidence for higher spatial frequencies in the source brightness distribution at all.

One possible explanation for this relation is that the sources are being scatter-broadened by turbulence in the ionized interstellar medium (ISM) of M31.
Essentially, each of our visibilities is a direct measurement of the correlation in the wave field on transverse separations of  between 100km and 8600km (our UV range).
It therefore measures the correlation in the plasma column density through M31 on these scales---equivalent to an angular resolution of 1--100\,picoarcseconds.

To determine whether this is a plausible explanation, we take the main components of the NE2001 model \citep{Cordes:2002}, which describes the ionized ISM of our Galaxy, and use this to predict, to an order of magnitude, the scattering within the inner part of M31.
Following \citet{Taylor:1993}, the scatter-broadened angle (in mas) for a source external to our Galaxy is given by
\begin{equation}
	\theta_{\mathrm{xgal}} = 128\,\mathit{SM}^{0.6}\,\nu^{-2.2} ,
	\label{eqn:scatter}
\end{equation}
where $\nu$ is the frequency in GHz and $\mathit{SM}$, the scattering measure is given by
\begin{equation}
	\mathit{SM}= \int^S n_e(S)^2\,F\,dS ,
	\label{eqn:dsm}
\end{equation}
where $S$ is the line of sight and $F$ is a scaling factor which relates the turbulence to the electron density squared.

\subsubsection{Scattering due to a galactic center component}
Observations show the existence of an extremely turbulent region close to Sgr~A*.
In NE2001 this region has a height and radius of $R_{GC}$=0.145\,kpc and $H_{GC}$=0.026\,kpc respectively.
Both the $n_e$ and $F$ are orders of magnitude higher than elsewhere in the galaxy.

At 1.6\,GHz, the scatter-broadened size of a background source seen through this region is of order 1\arcmin.
Therefore the detection of 37W142, at a projected distance only 0.25\,kpc from the core already places a constraint on the extent of such an extreme scattering region around M31*.

This result is consistent with the findings of \citet{Ciardullo:1988}, who probed the electron density of the inner region of M31 via S\II\ line ratios.
They found a drop of several orders of magnitude in electron density over a range 0 to 0.2\,kpc in radius.

\subsubsection{Scattering due to a `thin' disk}
Outside the core, the large-scale component of NE2001 with the highest scattering measure is the inner `thin' disk (with the `thick' disk and spiral arms being more diffuse and with much lower turbulence).
This disk is given an annular morphology, originally to match a feature seen in CO, but favored over a morphology derived from the distribution of ultra-compact H\II\ regions \citep{Cordes:2003}.

The electron density of this component varies by two orders of magnitude over the inner 5\,kpc of the Galaxy.
However taking the peak characteristic density we find that it could give rise to scatter broadening of $\sim$50\,mas at 1.6\,GHz.
The observed source sizes that we see are therefore consistent with a scattering measure similar to that inferred for the inner regions of our own galaxy.
Furthermore, it is possible that some of the scatter that we see for sources further than 5\arcmin\ from M31* is also due to scatter broadening.

While the circular Gaussian fit is very poorly constrained due to the low S/N, the source 37W150 does appear to be resolved (\fig~\ref{fig:models}).
If this is due to scatter broadening, rather than intrinsic source structure, it would be unsurprising given this source's proximity to a large H\II\ region.

\section{Discussion}
\label{sec:discussion}
\subsection{Fraction of sources detected}
\begin{figure*}[t]
	\centering
	\includegraphics[width=\textwidth]{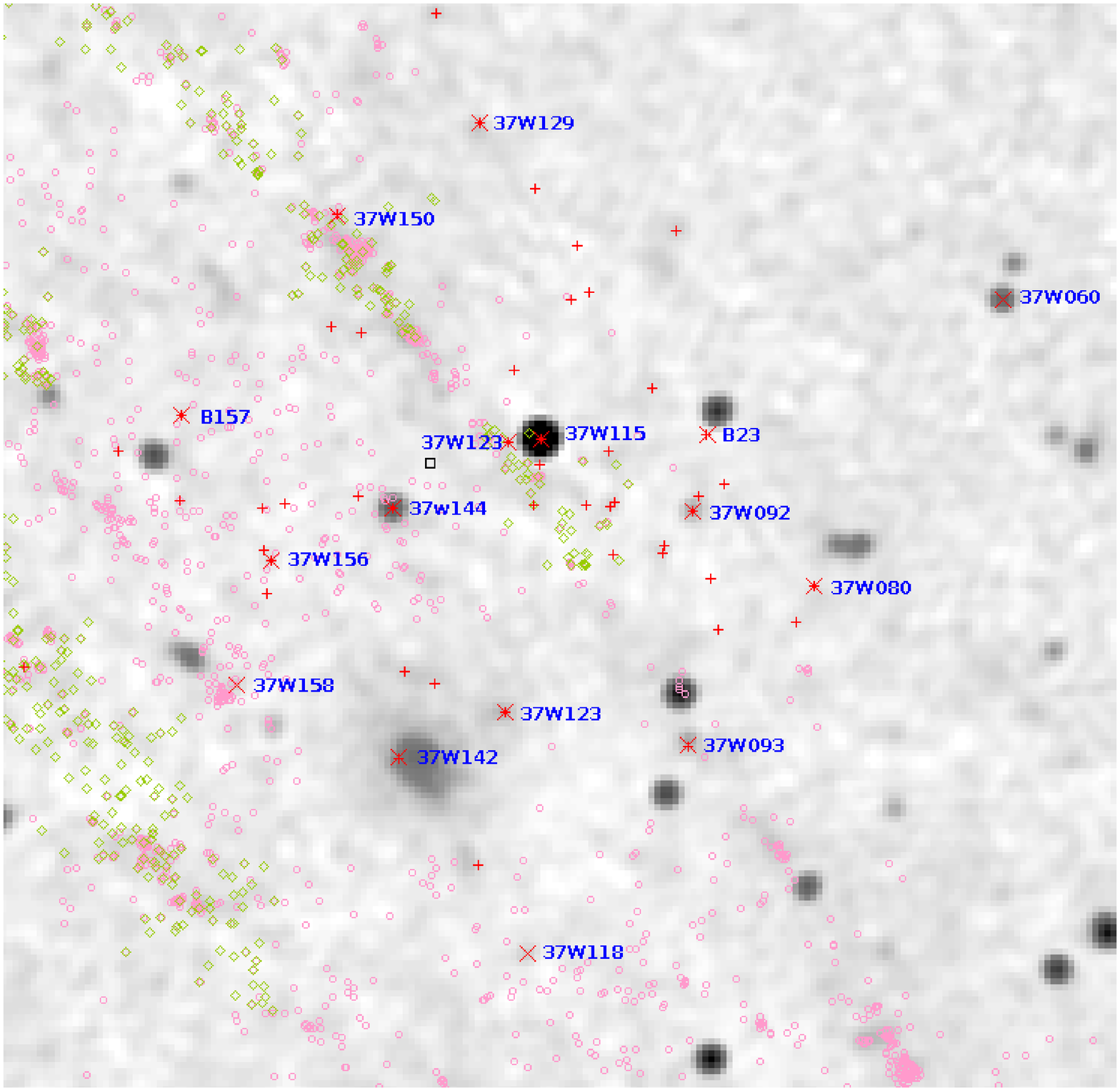}
	\figcaption{\label{fig:detected_undetected} Grayscale: image extracted from the NVSS survey \citep{nvss:1998}. Red 'x' with blue labels: detected sources. Red '+': unresolved B90 sources (major axis $>$ 6\arcsec) that would be detected at 6.6\,$\sigma$ were all their flux compact. Green diamonds: H\II\ regions from \citet{Walterbos:1992}. Pink circles: H\II\ regions from \citet{Azimlu:2011}.  Black square: pointing center.}
\end{figure*}

As discussed in \Sect~\ref{sec:results}, 16 individual sources were detected (see \fig~\ref{fig:detected_undetected}).
If we consider any field within 2.5\arcsec\ of another a duplicate, this gives us a detection rate of 16/217.
Recall, however, that no flux limit was included in our source selection criteria, for reasons set out in \sect~\ref{sec:correlation}.

For more realistic estimate detection we restrict ourselves to the B90 catalog and consider only those sources that would be detected at 6.6\,$\sigma$ or more (assuming no flux is resolved out).
This gives us a detection rate of 14/97.

Finally by considering sources that were unresolved or only very weakly resolved (major axis $<$6\arcsec), we find a detection rate of 13/48.
This is broadly consistent with the detection rate of \citet{Middelberg:2013}.

The 35/48 sources in this sample that were \emph{not} detected are also shown in \fig~\ref{fig:detected_undetected}.
There appear to be two double sources, a further three are associated with compact H\II\ regions and are therefore likely to be intrinsic to M31 and many appear to be clustered in a region close to the starforming ring.

Non-detection of a large number of objects in the starforming ring unresolved by the VLA indicate that a large number of sources intrinsic to M31 could be imaged by instruments intermediate in resolution between the VLA and the VLBA (such as long-baseline LOFAR, \emerlin\ and the SKA).

\subsection{Future Work}
The Galactic ionized interstellar medium and its turbulence are a subject of intense study.
Techniques used to probe the interstellar medium and the applications of an accurate model are enumerated in \citet{Cordes:2002}.

The difficulty of mapping the electron density distribution in different parts of the Galaxy from our position well within the plane should be noted.
For example, as recently as 1998 there was significant uncertainty over whether the region of enhanced scattering responsible for the observed angular size of Sgr~A* was at the Galactic center or somewhere in the plane along the line of sight  \citep{Lazio:1998}.

Utilizing the recent VLBA sensitivity upgrade and the techniques outlined here, it would be possible to probe in the ionized medium of \emph{another} galaxy, in detail, for the first time.
Furthermore, scatter broadening can be discerned from intrinsic size (which scales with $\lambda$ to the power of one or less) by its $\lambda^2$ dependence, allowing scatter-broadening to be quantified even in cases where there is small-scale structure.

We have scheduled multi-frequency \widefield\ VLBI observations with the aim of finding unambiguous evidence that the observed source sizes are due to scatter broadening; increasing the S/N on our detections; detecting a larger population of sources; and discriminating between scatter broadening and source structure for all sources.
This will also allow us to explore the apparent variability of some of the sources.

As the next generation of radio telescopes come online, scatter-broadening of background sources will become a remarkably powerful probe of the ionized ISM of other galaxies.
Simulations \citep{Wilman:2008} predict that at 1.4\,GHz there are approximately 2000 compact sources brighter than 10$\upmu$Jy per square degree.
Thus with only an order of magnitude sensitivity improvement it will be possible to probe the ISM of M31 and other nearby galaxies in exquisite detail.
Furthermore, the long-baseline SKA will also operate down to 500MHz meaning that it will be sensitive to scattering measures an order of magnitude weaker than those probed here.

The absolute positions of the 16 detected sources in M31 have been established with mas accuracy.
Relative to the in-beam calibrator source 37W144, however, the positional accuracy is much higher, in the range of several tens to several hundreds of microarcseconds, depending on angular separation and source brightness.
For most sources the S/N--limited fitting accuracy dominates the error budget, rather than systematic differential calibration errors.
In addition to 37W144, the sources 37W115, 37W158 and 37W092 are all bright and compact enough to serve as VLBI calibrators, and these 4 sources together could form a calibrator grid capable of supplying a relative reference frame accurate at the level of 10s of $\upmu$as for future astrometric studies of weak sources intrinsic to M31.

\subsection{Conclusions}
\begin{itemize}
	\item We have detected 16 compact sources across a region 35\arcmin\ in diameter.
	\item We have classified all 16 as background AGN based on their high compactness and compact and extended jet morphologies.
	\item The relative source positions that we have derived (with an accuracy $\sim$10\,$\upmu$as) can be used as the basis of a reference frame for any future astrometric studies of M31.
	\item Our source detection rate is broadly consistent with other \widefield\ VLBI studies. This rules out a large population of background objects scattered, absorbed or otherwise obscured by the intervening galaxy.
	\item We have probed the interstellar medium of another galaxy (M31), along several lines of sight, for the first time, and have found evidence of scatter broadening along the two lines of sight which pass closest to M31*.
	\item Upper limits on scatter broadening along the other lines of sight are broadly consistent with models for the ionized ISM of our own Galaxy.
	\item The detection of the source 37W142 places an upper limit of 0.25\,kpc on the radius of any extreme scattering region surrounding M31*.
	\item Measurements of scatter broadening of background sources at multiple GHz frequencies with wide-field VLBI has potential as a powerful probe of the ionized ISM of nearby galaxies.
\end{itemize}
We would like to thank Walter Brisken (NRAO) for assistance in data transfer.
We thank Stefanie Muehle and Bob Campbell (JIVE) for useful discussions regarding VLBI scheduling and correlation.
Extensive use was made of topcat \footnote{\url{http://www.starlink.ac.uk/topcat/}} \citep{Taylor:2005} and Aladin\footnote{\url{http://aladin.u-strasbg.fr/}} \citep{Bonnarel:2000}.
The National Radio Astronomy Observatory is a facility of the National Science Foundation operated under cooperative agreement by Associated Universities, Inc. 

\bibliographystyle{hapj}
\bibliography{mn-jour,tech}
\end{document}